\documentclass[12pt]{article}
\usepackage{pdfpages}
\usepackage{amsmath}
\usepackage{color}
\usepackage{graphicx}

\title{{\huge Free energy inference from partial work
    measurements in small systems}}

\author{M. Ribezzi-Crivellari \\ 
\\
Departament de Fisica Fonamental,\\
Universitat de Barcelona,\\
Diagonal 647, 08028 Barcelona, Spain\\
\\
and\\
\\
F. Ritort \\
\\
Departament de Fisica Fonamental,\\
Universitat de Barcelona,\\
Diagonal 647, 08028 Barcelona, Spain\\
\\
Ciber-BBN de Bioingeneria, Biomateriales y Nanomedicina,\\
Instituto de Salud Carlos III,\\
Madrid, Spain}

\date{}

\begin{document}
\maketitle
\newpage
\begin{abstract}
Fluctuation relations (FRs) are among the few existing general results in
non-equilibrium systems. Their verification requires the measurement of
the total work (or entropy production) performed on a system. 
Nevertheless in many cases only a partial measurement of the
work is possible. Here we consider FRs in
dual-trap optical tweezers where two different forces (one per
trap) are measured. With this setup we perform pulling experiments on
single molecules by moving one trap relative to the other. We
demonstrate that work should be measured using the force exerted by the
trap that is moved. The force that is measured in the trap at rest fails to provide 
the full dissipation in the system leading to a (incorrect) work definition that does not 
satisfy the FR. The implications to single-molecule experiments
and free energy measurements are discussed. In the case of symmetric
setups a new work definition, based on differential force measurements,
is introduced. This definition is best suited to measure free energies
as it shows faster convergence of estimators. We
discuss measurements using the (incorrect) work definition as an example
of partial work measurement. We show how to infer the full work distribution 
from the partial one via the FR. The inference
process does also yield quantitative information, e.g. the hydrodynamic
drag on the dumbbell. Results are also obtained for asymmetric dual-trap
setups. We suggest that this kind of inference could represent a new and
general application of FRs to extract information about 
irreversible processes in small systems.
\end{abstract}

\section*{Significance Statement}
Fluctuation Relations (FRs) provide general results about the full work
(or entropy production) distributions in non-equilibrium
systems. However, in many cases the full work is not measurable and only
partial work measurements are possible.  The latter do not fulfill a FR
and cannot be used to extract free energy differences from irreversible
work measurements. We propose a new application of FRs to infer the full
work distribution from partial work measurements. We prove this new type
of inference using dual-trap optical tweezers where two forces (one per
trap) are measured, allowing us to derive full and partial work
distributions. We derive a set of results of direct interest to single
molecule scientists and, more in general, to physicists and
biophysicists.

\section*{Introduction}
Fluctuation Relations (FRs) are mathematical equations connecting non
equilibrium work measurements to equilibium free energy differences. FRs, 
such as the Jarzynski Equality (JE) or the Crooks Fluctuation Relation (CFR)
have become a valuable tool in single-molecule biophysics where they are
used to measure folding free energies from irreversible pulling
experiments \cite{PNAS.hummer.2001,PNAS.hummer.2010}.  Such measurements
have been carried out with laser optical tweezers on different
nucleic acid structures such as hairpins 
\cite{liphardt2002equilibrium,nature.collin.2005,natphys.gupta.2011,natphys.alemany.2012},
G-quadruplexes \cite{dhakal2010coexistence,dhakal2013structural} and
proteins
\cite{cecconi2005direct,Nature.shank.2010,PNAS.gebhardt.2010,PNAS.yu.2012},
and with Atomic Force Microscopes on proteins \cite{prl.harris.2007} and
bi-molecular complexes \cite{PCCP.bizzarri.2010}.   
An important issue regarding FRs is the
correct definition of work, which rests on the correct identification of
configurational variables and control parameters. 
In the single-trap optical tweezers configuration this issue has been thoroughly discussed
\cite{douarche2005estimate,jchemphys.mossa.2009,alemany2011recent}.
\\
The situation about how to correctly measure work in small
systems becomes subtle when there are different forces applied to
the system. In this case theory gives the prescription to
correctly define the work ($W_{\Gamma}$) for a given trajectory ($\Gamma$): 
integrate the generalized force ($f_{\lambda}$)
(conjugated to the control parameter, $\lambda$) over $\lambda$ along $\Gamma$,
$W_{\Gamma}=\int_{\Gamma}f_{\lambda}d\lambda$. However, in some cases one
cannot measure the proper generalized force or has limited experimental access
to partial sources of the entropy production leading to what we call incorrect
or partial work measurements. A remarkable example of this situation are
dual-trap setups, mostly used as the high-resolution tool for single
molecule studies (Fig. \ref{Figure:opts}A). In this case a dumbbell formed by a
molecule tethered between two optically trapped beads is manipulated by
moving one trap relative to the other. In this setup two different forces (one per trap) can be measured
and at least two different work definitions are possible. In equilibrium
conditions, i.e. when the traps are not moved both forces are
equivalent: the forces acting on each bead have equal magnitude and
opposite sign. 
On the contrary, in pulling experiments, where
one trap is at rest (with respect to water) while the other is moved,
the two forces become inequivalent. This is so because the center of 
mass of the dumbbell drifts and the beads are affected by different
viscous drags (purple arrows in Fig. \ref{Figure:opts}B).
In such conditions theory prescribes that   
the full thermodynamic work must be defined on the force measured at the
moving trap whereas the force measured in the trap at rest (with respect to
water) leads to a {\it partial work measurement} which, as we show below, entails a systematic error in free energy
estimates. The difference between both works equals the dissipation by
the center of mass of the dumbbell, which is only correctly accounted for in the correct work definition. 
\\ In this paper we combine theory and experiments in a dual-trap
  setup to demonstrate several results.  First, we show that if the
  wrong work definition is used, free energy estimates will be
  flawed. The error is especially severe in the case of unidirectional
  work estimates (e.g. JE) while it influences bi-directional estimates
  to a lesser extent. This fact is not purely academical: measuring the
  force in the trap at rest is experimentally easier in dual-trap setups
  and, in fact, many groups choose to do so
  \cite{natphys.gupta.2011,nature.van.2008,mmb.cisse.2011}.  For example
  if different lasers are used for trapping and detection, measuring the
  force in the moving trap poses the additional challenge of keeping the
  trapping and detection lasers aligned while moving them.  Second, we
  show how it is possible, by using the CFR, to {\it infer} the
  the full work distribution from partial work
  measurements.  We demonstrate this new type of inference in our
  dual-trap setup by showing how to reconstruct the correct work
  distribution (i.e the one we would have measured in the moving optical
  trap) from partial work measurements in the wrong optical trap
  (i.e. the one at rest with respect to water) by using the CFR.  In
  particular, for symmetric setups the correct work distribution can be
  directly inferred by simply shifting the partial work distribution.
  In asymmetric setups inference is still possible in the framework of a
  Gaussian approximation, but the knowledge of some equilibrium
  properties of the system is still required. This type of inference should be seen an example of a more general application of
  FRs which aims at extracting information about the total entropy
  production of a nonequilibrium system from partial entropy production
  measurements. This allows us to determine the average power dissipated by the center of mass
  of the dumbbell, from which we can extract the corresponding
  hydrodynamic coefficient thereby avoiding direct hydrodynamic
  measurements. Moreover, we argue this type of inference might find
  applicability to future
  biophysical experiments where the sources of entropy production are
  not directly measurable, e.g. ATP-dependent motor translocation where
  the hydrolysis reaction cycle cannot be followed one ATP at a
  time. Finally we show how, in symmetric setups, the work definition
  satisfying the CFR is not unique. In particular the {\it differential
    work}, based on differential force measurements
  \cite{PNAS.moffitt.2006}, still satisfies the CFR and leads to the
  least biased free energy estimates. The distinguishing feature of this
  work definition is that it completely filters out the dissipation due
  to the motion of the center of mass of the dumbbell.  Throughout the
  paper, and for simplicity and pedagogical reasons, most of the
  derivations, calculations and experiments are shown for symmetric
  dual-trap setups whereas the asymmetric case is discussed towards the
  end of the paper.

\section*{The model} In dual-trap setups a molecule is stretched by two optical traps, the control parameter $\lambda$ being the trap-to-trap distance. 
Let A,B denote the two optical traps. When one trap (say trap A) is
moved with respect to the bath while trap B is at rest, suitable
configurational variables are the positions of the beads, both measured
from the center of the trap at rest (trap B).  We shall denote these
variables by $y_A$ and $y_B$, Fig. \ref{Figure:opts}B. The total energy
of the system is composed of three terms:
\begin{equation}\label{pot}
U(y_A,y_B,\lambda)=U_m(y_A-y_B)+\frac{k_B}{2}y_B^2+\frac{k_A}{2}\left(\lambda-y_A\right)^2\,\,\,,
\end{equation} 
where the quadratic terms model the potential of the optical trap and $U_m$ describes the properties of the tether. 
However one could measure the positions of the beads and the trap-to-trap distance in the moving frame of trap A ($x_A$, $x_B$ and $-\lambda$ in Fig. \ref{Figure:opts}B),
and the potential energy in Eq. \eqref{pot} would be written as:
\begin{equation}\label{p2}
U'(x_A,x_B,\lambda)=U_m(x_A-x_B)+\frac{k_A}{2}x_A^2+\frac{k_B}{2}\left(\lambda+x_B\right)^2\,\,\,,
\end{equation}
where $y_A-x_A=y_B-x_B=\lambda$ ($x_A$ and $x_B$ are negative).
Central to our analysis will be the equation connecting the potential $U$ and the work performed on the system by changing the control parameter \cite{PNAS.hummer.2001}:
\begin{equation}\label{w}
 W=\int_{t_i}^{t_f} dt \partial_t U\,.
\end{equation}
From Eq. \eqref{pot} we get:
\begin{equation}\label{trtt}
W=\int_{t_i}^{t_f} f_A\dot\lambda dt,
\end{equation}
with $f_A=-k_A(y_A-\lambda)$.
Inserting $U'$ instead of $U$ in \eqref{w} gives:
\begin{equation}\label{wsb}
W'=\int_{t_i}^{t_f} dt \partial_t U'=-\int_{t_i}^{t_f} f_B\dot \lambda dt,
\end{equation}
Despite of their similarity we will show that $W$ and $W'$ are remarkably different.
In fact, from the reference frame of trap A, the bath is seen to flow with velocity $-\dot
\lambda$ (Fig. \ref{Figure:opts}B). Because of this flow an experiment in which trap A
is moved is not Galilean equivalent to one in which trap B is moved.  In the presence of a flow the connection between potential
and work, Eq. \eqref{w}, is not valid anymore. In fact thermodynamic
work measurements must be based on the force measured in the trap being
moved. This fact has been discussed in \cite{prl.speck.2008}, yet the
implications to single-molecule experiments have never been pointed out.
Summarizing, if trap A is moved and B is at rest with respect to water,
using the work $W$ (Eq. \eqref{trtt}) in the JE leads to correct free
energy estimates whereas using $W'$ (Eq. \eqref{wsb}) in the JE leads to
a systematic error. Below we quantify such error in detail.
The difference between $W$ and $W'$ can be readily discussed in
symmetric setups ($k_A=k_B=k$) where calculations are much simpler.  To do this we switch to a new
coordinate system: $x_+=\frac{1}{2}(y_A+y_B),\,x_-=y_A-y_B$.  Here $x_+$
is the position of the geometric center of the dumbbell, while $x_-$ is
the differential coordinate \cite{PNAS.moffitt.2006}.  In this
coordinate system the potential Eq. \eqref{pot} reads
$U(x_-,x_+;\lambda)=U^-(x_-;\lambda)+U^+(x_+;\lambda)$:
\begin{equation}\label{potos}
\begin{split}
U^-(x_-,\lambda)&=U_m(x_-)+\frac{k}{4}\left(x_--\lambda\right)^2,\\U^+(x_+,\lambda)&=k\left(x_+-\frac{\lambda}{2}\right)^2.
\end{split}
\end{equation}
The potential energy term $U_+$ associated to $x_+$ is that of a moving trap, a problem that has been addressed both with experiments and theory \cite{mazonka1999exactly,wang2002experimental},  
while the potential energy term $U_-$ associated to $x_-$ corresponds to pulling experiments performed using a single trap and a fixed point.
The dumbbell is in contact with an isothermal bath where the equilibrium
state is described by the Boltzmann distribution and the corresponding
partition function (we assume a weak system-environment coupling,
 a situation satisfied in our experimental conditions, Section S2 in the SI).
Consequently, the two degrees of freedom are uncoupled and the total partition function for the system factorizes:
\begin{equation}\label{parti}
\begin{split}
Z(\lambda)&=Z^+(\lambda)Z^-(\lambda)\\ 
\qquad Z^\pm(\lambda)&=\int dx_\pm \exp\left(-\beta U^\pm\right)\,\,\, ,
\end{split}
\end{equation}
with $\beta=(K_BT)^{-1}$, $T$ being the temperature and $K_B$ being Boltzmann constant.
As a consequence free energy changes in the system can be decomposed into two contributions:
\begin{equation}\label{bl}
\Delta G=\Delta G^++\Delta G^-,\qquad \Delta G^{\pm}=-\beta^{-1}\log Z^\pm . 
\end{equation}
Work can also be decomposed into two contributions, each regarding one of the subsystems:
$ W^\pm=\int \partial_t U^\pm(x_\pm;\lambda)dt.$
Here $W^-$ contains the work done in stretching the molecule while $W^+$ is pure dissipation due to the movement of the center of mass of the dumbbell.
Note that:
\begin{equation}\label{wdecompo}
W=W^-+W^+\,\,, \qquad W'=W^--W^+\,\,,
\end{equation}
which shows that the difference between $W$ and $W'$ is entirely due to $W^+$.
The JE holds for $W$, the standard work definition, so that 
\begin{equation}\label{jazzy}
\Delta G=-\beta^{-1}\log\langle \exp{\left(-\beta W\right)}\rangle.
\end{equation}
Inserting Eqs. \eqref{bl},\eqref{wdecompo} in \eqref{jazzy} we get:
$\Delta G^++\Delta G^-=-\beta^{-1} \langle \exp{\left(-\beta
  \left(W^++W^-\right)\right)}\rangle$. In symmetric setups $W^+$ and
$W^-$ are independently distributed random variables (see Sections
  S1 and S3 in the SI) and we can conclude
that:
\begin{equation}\label{jzpm}
\Delta G^\pm=-\beta^{-1}\log\langle e^{-\beta W^\pm}\rangle\qquad.
\end{equation}
Using the JE on both $W$ and $W'$ we get two different free energy estimates: $\Delta G$ (Eq. \eqref{jazzy}) and
$\Delta G'=-\beta^{-1} \log\langle \exp\left(-\beta W'\right)\rangle$. The error $\mathcal{E}$ committed by using $W'$ instead of $W$  can be quantified as:
\begin{equation}\label{bias}
\mathcal{E}=\Delta G-\Delta G'=-\beta^{-1}\log\frac{\langle\exp(-\beta W)\rangle}{\langle\exp(-\beta W')\rangle}.
\end{equation}
From Eqs. \eqref{wdecompo},\eqref{jazzy} and again using the fact $W^+$ and $W^-$ are independently distributed random variables we get $\langle \exp(-\beta W)\rangle=\langle \exp(-\beta \left(W^++W^-\right))\rangle=\langle \exp(-\beta W^+)\rangle\langle \exp(-\beta W^-)\rangle$ and
similarly for $W'$. As a consequence $\frac{\langle\exp(-\beta W)\rangle}{\langle\exp(-\beta W')\rangle}=\frac{\langle\exp(-\beta W^+)\rangle}{\langle\exp(+\beta W^+)\rangle}$ and Eq. \eqref{bias} is reduced to 
\begin{equation}\label{biasp}
\mathcal{E}= -\beta^{-1}\log\frac{\langle\exp(-\beta W^+)\rangle}{\langle\exp(+\beta W^+)\rangle}.
\end{equation}
Since $x_+$ is subject to a quadratic potential (Eq. \eqref{potos}), we expect $W^+$ to be a Gaussian random variable. This is not true in general for $W^-$ given that $x_-$ feels the nonlinear term $U_m$.
For Gaussian Random Variables we have:
\begin{equation}\label{gigi}
\beta^{-1}\log\langle \exp(\pm\beta W^+)\rangle=\langle W^+\rangle\pm\frac{\beta}{2}\sigma^2_+, 
\end{equation}
where by $\sigma_+^2$ we denote the variance of $W^+$.  Moreover 
dragging a trapped bead in a fluid causes no free energy change, so that:
\begin{equation}\label{gromp}
\Delta G^+=-\beta^{-1} \log\langle \exp(-\beta W^+)\rangle=\langle W^+ \rangle-\frac{\beta}{2}\sigma_+^2=0 
\end{equation}
or $\langle W^+ \rangle=\frac{\beta}{2}\sigma_+^2$. 
Inserting Eqs. \eqref{gigi} and \eqref{gromp} in Eq. \eqref{biasp} we get:
\begin{equation}\label{ppo}
\mathcal{E}=\langle W^+\rangle+\frac{\beta}{2}\sigma_+^2=2\langle W^+\rangle.
\end{equation}
Equation \eqref{ppo} gives $\mathcal{E}=\beta\sigma_+^2>0$ showing that $\Delta G'$ is lower than $\Delta G$ (Eq. \eqref{bias}). Interestingly enough, using $W'$ instead of $W$ in the JE leads
to free energy estimates in apparent violation of the second law.
The error on free energy estimates obtained using $W'$ instead of $W$ is proportional to the mean work performed on the center of the dumbbell.
This mean work $\langle W^+\rangle$ is just the mean friction force times the total trap displacement $\Delta \lambda$:
\begin{equation}\label{defiw}
\langle W^+\rangle=\gamma_+\frac{\dot \lambda}{2} \Delta \lambda\qquad,
\end{equation}
where $\gamma_+$ is the friction coefficient of the drag force opposing
the movement of the geometric center of the dumbbell. The value of
$\gamma_+$ can be independently obtained from equilibrium measurements \cite{Prl.meiners.1999} (Sections S4,S5 in the SI).

\section*{Differential Work Measurements}

Equations \eqref{wdecompo} and \eqref{gromp} show that free energy estimates based on the standard work $W$ and the differential work $W^-$ are equivalent:
\begin{equation}\label{ddcc}
 \begin{split}
\Delta G&=-\beta^{-1}\log\langle \exp(-\beta W)\rangle=\\
&= -\beta^{-1}\log\langle \exp(-\beta W^-)\rangle-\beta^{-1}\log\langle \exp(-\beta W^+)\rangle=\\
&=-\beta^{-1}\log\langle \exp(-\beta W^-)\rangle=\Delta G^{-}.
\end{split}
\end{equation}
We stress that this is only true for symmetric setups where $W^+$ and
$W^-$ are independent random variables. The case of asymmetric setups
is discussed further below. Therefore $W^-$ can be used for free energy determination, as it has been done in \cite{PNAS.gebhardt.2010}, although without discussion.
Equation \eqref{ddcc} does only hold when the number of work measurements, $N$, tends to infinity. In all practical cases we deal with finite $N$ and the Jarzynski estimator is biased \cite{PNAS.gore.2003,prl.palassini.2011}. The bias is strongly 
linked to the typical dissipation $D_\text{typ}$ and a reliable estimate of free energy differences requires a number of work measurements which scales as $N\simeq \exp\left(D_\text{typ}\right)$ \cite{ritort2002two,PrE.jarzynski.2006},
so that even a small reduction in $D_\text{typ}$ entails a considerable improvement in the convergence of free energy estimators. Moreover the bias is superadditive. Let us consider for simplicity 
Gaussian Work Distributions.  In this case the bias, $B^\text{GWD}_N$, in the large $N$ limit is a function of the variance of the distribution $\sigma^2$ and of $N$ \cite{PNAS.gore.2003}:
\begin{equation}\label{fgf}
B^\text{GWD}_N=\frac{\exp{(\beta^2\sigma^2}-1)}{2\beta N}\qquad .
\end{equation}
$B^\text{GWD}_N$ is a convex function of $\sigma$, and is superadditive i.e. $B^\text{GWD}_N(\sigma^2+\phi^2)>B^\text{GWD}_N(\sigma^2)+B^\text{GWD}_N(\phi^2)$. 
This means that, should the work be the sum of two independent Gaussian contributions, the bias on the sum is greater than the sum of the biases. 
Although Eq. \eqref{fgf} was derived under strong assumptions, superadditivity does also hold for other theoretical expressions for the bias and has been checked in our experimental data (see below). 
Let us introduce the following Jarzynski estimators for finite $N$:
\begin{eqnarray}
\Delta G_N=-\beta^{-1}\log \frac{1}{N}\sum_{i=1}^Ne^{-\beta W_i},\label{e1}\\
\Delta G_N^\pm=-\beta^{-1}\log \frac{1}{N}\sum_{i=1}^Ne^{-\beta W_i^\pm}\label{e2}
\end{eqnarray}
and the corresponding bias functions:
\begin{eqnarray}
B_N&=&\Delta G_N-\Delta G \label{fffd}\\
B^{\pm}_N&=&\Delta G^\pm_N-\Delta G.\label{fffe}
\end{eqnarray}
Since $W=W^++W^-$, superadditivity guarantees:
\begin{equation}\label{hiera}
B_N\geq B^{-}_N+B^{+}_N\geq  B^{-}_N.
\end{equation}
Because of Eq. \eqref{hiera} differential work measurements always improve the convergence of free energy estimates in dual-trap setups. 
This is especially important in all those cases in which bidirectional methods (e.g. the CFR) cannot be used and one has 
to employ unidirectional methods.

\section*{Pulling on ds-DNA} The theory discussed so far has been put to test in a series of pulling experiments performed in a recently developed dual-trap optical tweezers setup which directly measures force in each trap \cite{ribezzi2012force,ribezzi2013counter}.
The setup can move the two optical traps independently and measure their relative position with sub-nanometer accuracy, giving direct access to both $W$ and $W'$. 
In these experiments 3 kb ds-DNA tethers ($\simeq1$ $\mu$m in contour
length) were stretched between 1 and 3 pN (Fig. \ref{Figure:opts}C)
in a symmetric dual-trap
setup ($k_A=k_B=0.02$pN/nm) using 4 $\mu$m silica beads as force probes.  The experiments were performed moving one of the two traps (trap A) with respect to the lab frame 
and leaving trap B at rest. All experiments were performed in PBS buffer (pH 7.4, 1M NaCl).
We chose cyclical protocols ($\lambda_t:\, \lambda_0=\lambda_{T_{c}}$, where $T_{c}$ is the total duration of the cyclic protocol). The excursion of the control parameter, $\Delta \lambda=\lambda_{T_{c}/2}-\lambda_0$,  
was varied between 200, 400 and 600 nm, while the pulling speed was varied between 1.35$\pm$0.05, 4.3$\pm$0.1 and 7.2$\pm$1 $\mu$m/s.  Given the force-distance curves the total dissipation along cycles was measured:
\begin{equation}\label{dd}
D=\oint d\lambda f_A\,\,, \qquad D'=-\oint d\lambda f_B\,\,. 
\end{equation}
The CFR \cite{crooks1999entropy} is a symmetry relation between the work distribution associated to the forward ($P_F$) and time reversed ($P_R$) protocols:
\begin{equation}\label{crozza}
P_F(W)=P_R(-W)\exp(\beta\left(W-\Delta G\right)).
\end{equation}
In the case of cyclic protocols $P_F=P_R=P$ and $\Delta G=0$ so that the CFR takes the form:
\begin{equation}\label{cfr}
P(D)=\exp{\left(\beta D\right)}P(-D)\qquad.
\end{equation}
Such symmetry of the probability distribution for $D$ can be directly tested in cases where negative dissipation events are observed. In Fig. \ref{Figure:multi}A we show measured work histograms (solid points, left hand side of Eq. \eqref{cfr}) 
and reconstructed histograms (open points, right hand side of Eq. \eqref{cfr}). If Eq. \eqref{cfr} is fulfilled then the measured and reconstructed histograms match each other. A quantitative measure of the deviation from Eq. \eqref{cfr} can be obtained
from the ratio $P(D)/P(-D)$, as shown in Fig. \ref{Figure:multi}B. Experimental data shows that $D$ fulfills the FR whereas $D'$ does not. 
In Fig. \ref{Figure:multi}C we show that the pdfs of $D^+$ and $D^-$, with:
\begin{eqnarray}
D^-=\frac{D+D'}{2}\label{oro}\\
D^+=\frac{D-D'}{2}\label{sboro},
\end{eqnarray}
are experimentally found to satisfy a FR as in Eq. \eqref{cfr}. $D^-$ is
just the differential work, $W^-$, Eq. \eqref{wdecompo} evaluated on a
cyclic protocol, whereas $D^+$ is the dissipation due to the movement of
the center of mass of the dumbbell.  Summarizing, although in general
$W$ is the only observable we expect to fulfill a FR, in symmetric
setups two new FRs emerge, for $W^-$ and $W^+$.  In Fig. \ref{Figure:seifert}A,B we
compare the predictions of Eqs. \eqref{bias}, \eqref{defiw} with
experimental results for different pulling speeds, $\dot \lambda$, and
different displacements $\Delta \lambda$.  Equation \eqref{defiw} must
be used to correct free energy estimates obtained in all those dual-trap
setups which do not measure the force applied by the trap which is being
moved (as in \cite{natphys.gupta.2011,nature.van.2008,mmb.cisse.2011}).
The advantages of using $W^-$ in free energy estimates are shown in
Fig. \ref{Figure:seifert}C.  There we show the convergence of the
Jarzynski estimator with sample size for the cycles in
Fig. \ref{Figure:opts}C. Being evaluated over cycles, the expected free
energy change is zero.  The convergence of the estimator is faster for
$W^-$ than for $W$ in the three cases (Eq. \eqref{hiera}). The effect is
enhanced in our experiments by the high pulling speed (in the range 1-7
$\mu$m) and by the large bead radius (2 $\mu$m).  Let us note that due
to the finite lifetime of molecular tethers and unavoidable drift
effects, raising the pulling speed is a convenient strategy to improve
the quality of free energy estimates. Similar results have
been found also at low pulling speeds where, again, $W'$ does not
satisfy the CFR (Section S7 in the SI).

\section*{Experiments on DNA hairpins} 
Fluctuation theorems are used to extract folding free energies for nucleic acid secondary structures or proteins. 
We further tested the different work definitions by performing pulling experiments on a 20bp DNA hairpin (Fig. 4A) at a $0.96\pm0.02$ $\mu$m/s pulling speed in the same dual-trap
setup as in the previous dsDNA experiments. In this case the work performed during the unfolding and refolding of the molecule were 
considered separately, as it is customary for free energy determination. In Fig. 4B we present forward and reverse work histograms for $W,W'$ and $W^+$. Again $W$ 
and $W^-$ both fulfill the CFR  but $W$ shows higher dissipation than $W^-$, resulting in slower convergence of unidirectional free energy estimators (Fig. 4C). 
The difference between unidirectional free energy estimates
based on $W$ and $W^-$ is in this case $\simeq 1$ $K_BT$. 
As previously discussed for double-stranded DNA (Fig. 2A), $W'$ does not fulfill the CFR and,
as a consequence, unidirectional free energy estimates based on $W'$ are flawed. In our experimental conditions the error committed by using the wrong work definition 
is again positive and equal to $\mathcal{E}\simeq3$ $K_B T$. As previously discussed this leads to a negative average dissipated work, apparently violating the second law. 
It must be noted that the difference in free energy estimates based on $W^-$ and $W$ is a finite-size effect, whose magnitude decreases when an increasing number of work 
measurements is considered. On the contrary 
the error committed by using $W'$ does not vanish by increasing the number of work measurements. It can be noted from Fig 4B that, although $W'$ does not fulfill the 
CFR and gives wrong unidirectional estimates, its forward and reverse
distributions apparently cross at $W'=\Delta G$ within the experimental
error. Although this could be used for free energy determination
  the result should be taken with caution as we have no general proof
that this should happen in all cases.

\section*{Free energy inference from partial work measurements}

The JE and CFR are statements on the statistics of the total dissipation
in irreversible thermodynamic transformations.  The need to measure the
total dissipation limits the range of applicability of these and other FRs. 
For example testing FRs concerning the dynamics of
molecular motors would need the simultaneous measurement of both the
work performed by the motor and the number of hydrolyzed ATPs. Here we
demonstrate that, at least in some cases, a different approach is
possible. Let us start by considering the simple case of symmetric
setups as developed in the previous sections. In the experiments
discussed so far there are two sources of dissipation that we
were able to measure and characterize separately: the motion of the
dumbbell and the dissipation of the differential coordinate.We
already learnt that $W$ satisfies a FR while $W'$ does
not. Moreover we know that $W=(W_-+W_+)/2$ and $W'=(W_--W_+)/2$, and
that, being $W^+$ and $W^-$ uncorrelated random variables, $W$ and
$W'$ have the same variance. Imagine now to have only partial
information on the system.  For example, one could be able to measure
force only in the trap at rest, as many experimental setups do. With
this information, and in absence of any guiding principle, no statement
about the total entropy production is possible. We will find such
guiding principle if we assume the FR to hold for $W$.  Knowing
 that $W$ and $W'$ have the same variance, we just have to shift
the work distribution $P'(W')$ by $W=W'+\Delta$ to get a new
distribution that satisfies the CFR.  In practice this is done starting
from the set of $W'$ values and tuning the value of $\Delta$
(Fig. \ref{Figure5}A) until $P(W)=P'(W-\Delta)$ fulfills the CFR
(Fig. \ref{Figure5}B).  In the case of the hairpin, this same shifting
procedure is operated for both forward and reverse work
distributions. Again the value of $\Delta$ is tuned
(Fig. \ref{Figure5}C) until the CFR symmetry is recovered
(Fig. \ref{Figure5}D). The unique value of $\Delta$ that restores the
validity of the CFR equals the average work dissipated by the motion of
center of mass of the dumbbell, giving the hydrodynamic coefficient
$\gamma_+$ via Eq. \eqref{defiw}. Let us note that, once the work
distribution in the correct trap (i.e. the moving trap) has been
recovered, then we could also extract the correct free energy difference
(the value of $\Delta G$, Fig. \ref{Figure5}E) and infer the
distribution for the differential work $W_-$ by deconvolution.
\\
The extension of this analysis to the asymmetric case is more complex
but equally interesting (Fig. 6A). The decomposition of $W,W'$ in $W^+$ and $W^-$ (Eq. \eqref{wdecompo}) is
still possible although $W^+,W^-$ are not uncorrelated variables
anymore and neither $W^+$ nor $W^-$ satisfy a FR. In this
general case only $W$ satisfies a FR (but not $W'$, $W^+$,
$W^-$). Remarkably enough, in the framework of a Gaussian approximation, it is still possible to infer the correct work
distribution $P(W)$ out of partial work $W'$ measurements. The analysis
is presented in Section S6 of the SI. In this case it is enough to know the trap and molecular
stiffnesses $k_A,k_B,k_m$ i.e. some equilibrium properties of the system,
for a successful inference. To reconstruct $P(W)$ both the mean and
the variance of $P'(W')$ must be changed, which can be achieved by doing
a convolution between the $P'(W')$ and a normal distribution:
$P_{\Delta,\Sigma} =P'\star {\cal N}(\Delta,\Sigma)$ where
$\star$ denotes the convolution operator and ${\cal
  N}(\Delta,\Sigma)$ is a normal distribution with mean $\Delta$ and
standard deviation $\Sigma$. Starting from a distribution $P'(W')$ there
are infinitely many choices of $\Delta$ and $\Sigma$ which yield a
$P_{\Delta,\Sigma}(W)$ satisfying the CFR. Indeed, let us suppose that the
pair $\Delta^*,\Sigma^*$ is such that $P_{\Delta^*,\Sigma^*}(W)$
satisfies the CFR. Then it is easy to check that
$P_{\Delta^*+\phi,\sqrt{\Sigma^{*2}+2\phi K_BT}}$ will also satisfy the CFR
for any $\phi$ (Fig. 6B). In this situation the inference cannot rest on
the CFR alone. Explicit calculations in the Gaussian
case (section S6 in SI) show that variances ($\sigma^2$) and means
($\langle...\rangle$) of $P(W)$ and $P'(W')$ are related by an Asymmetry Factor (AF),
\begin{equation}
AF(k_A,k_m,k_B)=\frac{\sigma_W^2-\sigma_{W'}^2}{\langle W\rangle-\langle
  W'\rangle}=K_BT \frac{4k_m(k_A-k_B)}{k_A(k_B+2k_m)}
\label{eqAF}
\end{equation}
which only depends on {\it equilibrium properties} such as the stiffnesses of 
the different elements (section S6.3 in the SI). Knowing the AF allows us to select the unique pair
$\Delta,\Sigma$ such that $AF= \Sigma^2/\Delta$ with
$P_{\Delta,\Sigma}(W)$ satisfying the CFR. The inference procedure
can be described with a very simple formula (Section S6 in the SI). The key idea is to proceed as previously done in the case of
symmetric setups by just shifting the mean of $P'(W')$ by a parameter $\delta$
until the CFR is satisfied, i.e. $\Delta=\delta,\Sigma =0$. From the
values of $AF$ and $\delta$ we can reconstruct $P(W)$ by using the formulae,
\begin{equation}
\Delta=\frac{2\delta}{2-\beta AF}\,\,;\,\,\Sigma^2=\frac{2\delta AF}{2-\beta AF}
\label{eqdeltasigma}
\end{equation}
The inference procedure for an asymmetric setup is shown in Fig. 6C,6D
for cyclic ds-DNA pulling experiments.  For non-cyclic pulls ($\Delta
G\ne 0$) the procedure can be easily generalized in the line of what has
been shown for the case of hairpin in the symmetric setup (Fig. 5C).

\section*{Discussion}
%

FRs are among the few general exact results
in non-equilibrium statistical mechanics. Their validity has been
already extensively tested in different systems, ranging from single
molecules to single electron transistors, and in different conditions
(steady state dynamics, irreversible transformations between steady
states, transient nonequilibrium states). At the present stage, the main
widespread application of FR is free energy recovery from
non-equilibrium pulling experiments in the single molecule field. What
we are presenting here is a new application of FR for
inference. All FRs are statements about the statistics of the total
entropy production in a system plus the environment. If some part of
the entropy production is missed or inadequately considered FRs will
in general not hold. This is why, for irreversible transformations
between equilibrium states, we have a FR for the dissipated work
(which is the total entropy production) but not for the dissipated
heat (which is just the entropy production in the environment). The
main tenet is now that the violation of FRs in a given setting
provides useful information: it is an evidence that some
contribution to the total entropy production is being missed.
We have given rigorous examples in which the violation
of FRs can be used to characterize the missing entropy production.
Remarkably, in our model system, one could even replace the moving trap by a moving
micropipette, an object lacking any measurement capability, and still infer the
work distribution exerted by that object on the
molecular system (this extremely asymmetric setup would still be
described by Eq.(\ref{eqAF}), with $k_A\to\infty$ and $AF=4K_BTk_m/(k_B+2k_m)$).
These results open the exciting prospect of extending and applying these ideas 
to steady state systems, such as molecular motors, to extract useful information about their mechanochemical cycle.

\section*{Conclusions} 
In order to give it a clear and definite meaning to free energy inference we have discussed irreversible 
transformations between equilibrium states performed with dual-trap optical tweezers. In these
experiments a molecular tether is attached between two beads
which are manipulated with two optical traps. The irreversible
transformation is performed by increasing the trap-to-trap distance at
a finite speed. In this kind of transformations the dissipated work
equals the total entropy production leading to our first
result: in pulling experiments work $W$ must be defined on
the force measured in the trap which is moved with respect to the
thermal bath. The force measured in the trap at rest gives rise to a
work definition, $W'$, which does not satisfy the FR and is unsuitable to
extract free energy differences. We have called $W'$ a partial work measurement
because it misses part of the total dissipation. This result is of
direct interest to experimentalists: many optical tweezers setups are designed
so that they can only measure $W'$. We have thus imagined a situation in which 
$W'$ is measurable while $W$ is not and asked the question: can we infer the distribution of $W$ 
from that of $W'$? If the question is asked in full generality, without any system-specific
information, the answer is probably negative. Knowing only the extent of
violation of the FR will be of little use, in general some additional system specific information will 
be needed for a successful inference. Here we discussed free energy inference in the framework of a Gaussian
approximation, the extent to which such inference is generally possible should be the subject of future studies.
Let us summarize our main results:
\begin{itemize}
\item A symmetry of the system can be crucial for the inference. For
  left-right symmetric systems (as exemplified in our symmetric
  dual-trap setup) the $P(W)$ can be inferred from $P'(W')$ just by
  imposing that the former satisfies the CFR. When symmetry
  considerations cannot be used, the knowledge of some equilibrium
  properties of the system may suffice to successfully guide the
  inference (such as the stiffnesses for the asymmetric setup).

\item The inference process can be used both to recover the full
  dissipation spectrum plus additional information about the {\em
    hidden} entropy source. In our specific dual-trap example $W'$ does
  not account for the dissipation due to the movement of the
  center-of-mass of the dumbbell and the inference procedure can be seen
  as a method to measure the associated hydrodynamic drag.

\item We stress the benefits of using symmetric dumbbells
   in single molecule manipulation. In this case an alternative work definition, the
{\em differential work} $W^-$, fulfills the CFR and is thus suitable for free energy
measurements. Being $W^-$ less influenced by dissipation than $W$,  switching from $W$ to $W^-$ ensures faster
convergence of unidirectional free energy estimates.  For asymmetric
setups $W^-$ does not satisfy a FR anymore (only $W$ does) and cannot be used to extract free
energy differences.

\end{itemize}

A deep understanding of how to correctly define and measure
thermodynamic work in small systems (a long debated question in the past
20 years) is not just a fine detail for experimentalists and theorists
working in the single molecule field, but an essential question
pertaining to all areas of modern science interested in energy transfer
processes at the nanoscale. The new added feature of free energy inference
discovered in this paper paves the way to apply FRs to new problems and
contexts. This remains among the most interesting open problems in this exciting field.

\section*{Methods}

{\bf Buffers and DNA substrates.} All experiments were
  performed in PBS Buffer 1M NaCl at 25$^\circ C$; 1 mg/ml BSA was added to
  passivate the surfaces and avoid nonspecific interactions. The dsDNA
  tether was obtained ligating a 1kb segment to a biotin-labeled
  oligo at one end and a dig-labeled oligo at the other end.  The DNA
  hairpin used in the experiments has short (20bp) molecular handles and
  was synthesized by hybridization and ligating three different
  oligos. One oligo is biotin-labeled and a second is
  dig-labeled. Details of the systhesis procedure are given in
  \cite{forns2011hand}.

{\bf Optical Tweezers Assay.} Measurements were
  performed with a highly stable miniaturized laser tweezers in the dual
  trap mode \cite{ribezzi2013counter}.  This instrument directly
  measures forces by linear momentum conservation. In all experiments we
  used silica beads with 4 $\mu$m diameter, which give a maximum
  trapping force around $20$ pN. Data is acquired at 1 kHz.

\section*{Acknowledgements}
We thank A. Alemany and M. Palassini for a critical reading of the manuscript. FR is supported by ICREA Academia 2008 grant. The research leading to these results 
has received funding from the European Union Seventh Framework Programme (FP7/2007-2013) under grant agreement $\text{n}^\text{o}$ 308850 INFERNOS.





\section*{Caption of Figure 1}

\begin{figure}[ht]
 \centering
 \includegraphics[width=.70\textwidth]{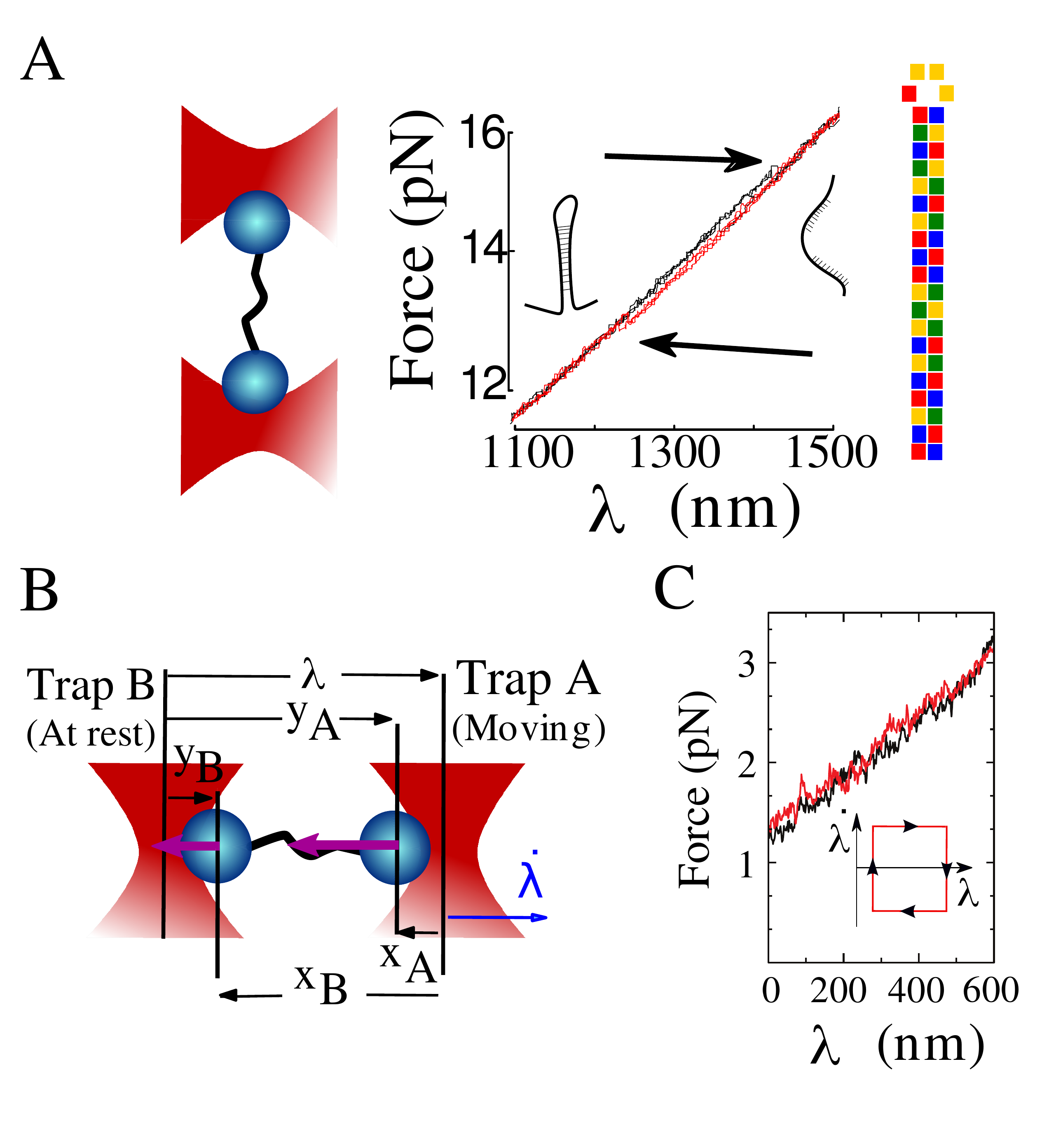}
\caption{\label{Figure:opts}}
\end{figure}

{\bf Pulling experiments with dual-trap optical tweezers.} A) Force-distance curves
in a pulling experiment on a 20 bp hairpin with our dual-trap setup. A molecular tether is attached
between two trapped beads. By increasing the distance, $\lambda$,
between the traps the tether is stretched or released until some
thermally activated reaction is triggered, e.g. the unfolding or folding
of a DNA hairpin, and detected as a force jump (black
arrows). The small force jump (0.2 pN) is due to the low-trap stiffness
 of our dual-trap setup ($\simeq$0.02 pN/nm). Inset: scheme of the
hairpin with color-coded sequence (A/T: yellow/green, G/C: red/blue). B)
Pulling experiments in a dual-trap setup where trap A is moved at speed
$\dot\lambda$ and trap B is at rest with respect to water. $\lambda$ is
the control parameter, $y_A$ and $y_B$ are the configurational
variables with respect to the moving trap A while $x_A$ and $x_B$ are the configurational
variables with respect to the trap at rest (trap B). C)
pulling curves (red stretching, black releasing) for a 3kb ds-DNA
tether in a dual-trap setup. Inset: the cyclic pulling protocol used in
the experiments.

%

\section*{Caption of Figure 2}

 \begin{figure}[h]
\centering
 \includegraphics[width=.70\textwidth]{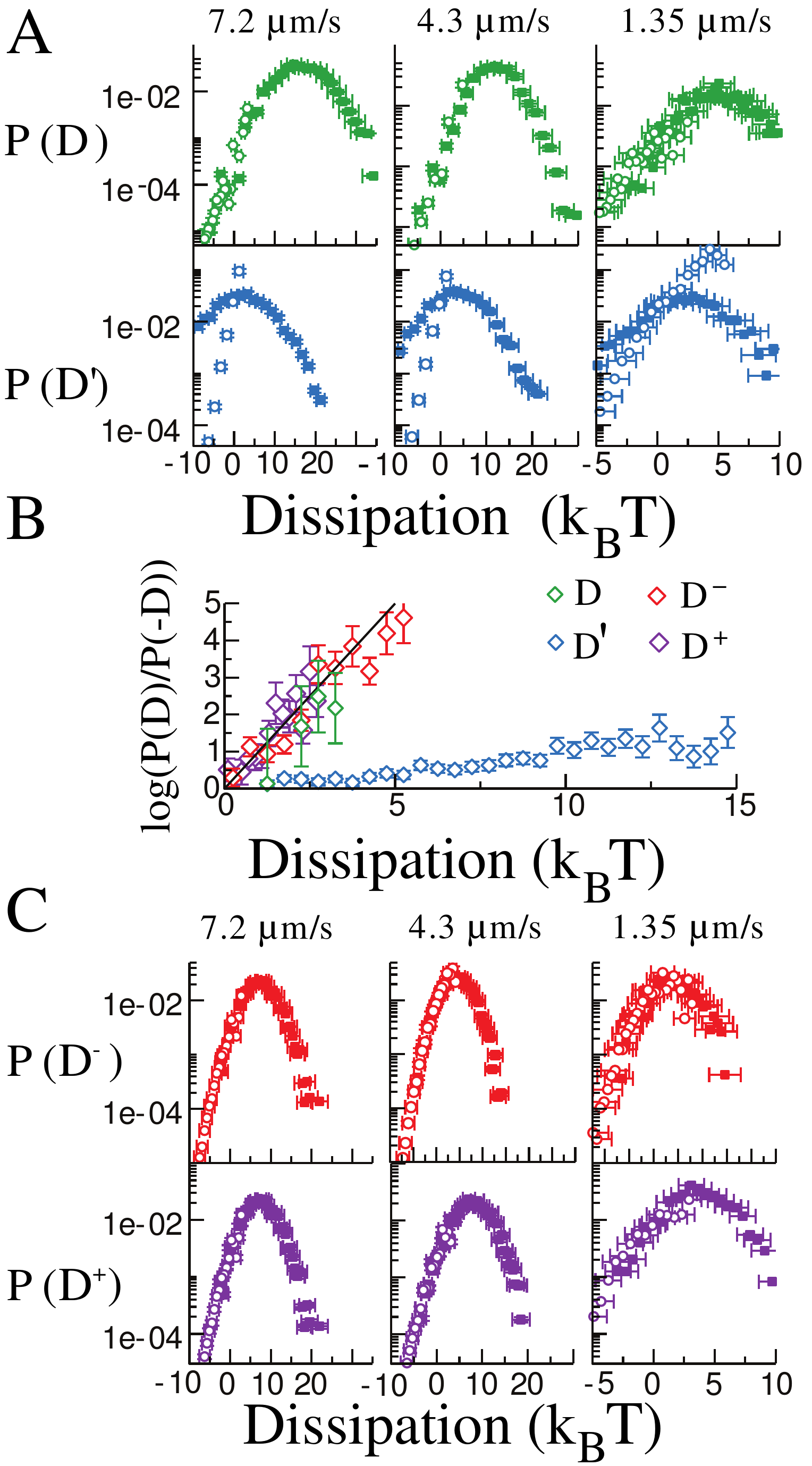}
 \caption{\label{Figure:multi}}
 \end{figure}

{\bf Work Distributions.} Work statistics obtained on cyclic protocols
with a 200 nm excursion with different pulling speeds (columns). Four
different work observables are considered. In each case the solid points
are direct work measurements (l.h.s. of Eq. \eqref{cfr}).  In order to
improve statistics these distributions are calculated as the convolution
of the distribution of the work performed while stretching with that
performed while releasing. For that we took all forward
  and reverse work values $W_F,W_R$ and combined them $W=W_F+W_R$ in
  order to get a joint work distribution for the cycle. Open symbols
are the reconstructed histogram (r.h.s. of Eq. \eqref{cfr}). Different
columns refer to different pulling speeds,$\dot \lambda$ , as shown on
top.  A) Comparison of the measured and reconstructed distributions
according to the two definitions of Eq. \eqref{dd} ($D,D'$).  The
distribution for $D$ satisfies the CFR Eq.\eqref{cfr}, i.e. the measured
and reconstructed distributions superimpose. The distribution for $D'$
does not satisfy it. Horizontal error bars represent the systematic
error in work measurements, while vertical error bars denote statistical
errors.  B) The CFR, Eq. \eqref{cfr} is satisfied within the
experimental error for $D,D^+$ and $D^-$ but not for $D'$. C) Comparison
between the measured and reconstructed distributions for $D^-$ and $D^+$
(Eq.\eqref{oro} and Eq. \eqref{sboro}) both of which satisfy the
CFR. Horizontal error bars represent the systematic error in work
measurements, while vertical error bars denote statistical errors.

\section*{Caption for Figure 3}

%
\begin{figure}[h]
  \centering
 \includegraphics[width=0.6\textwidth]{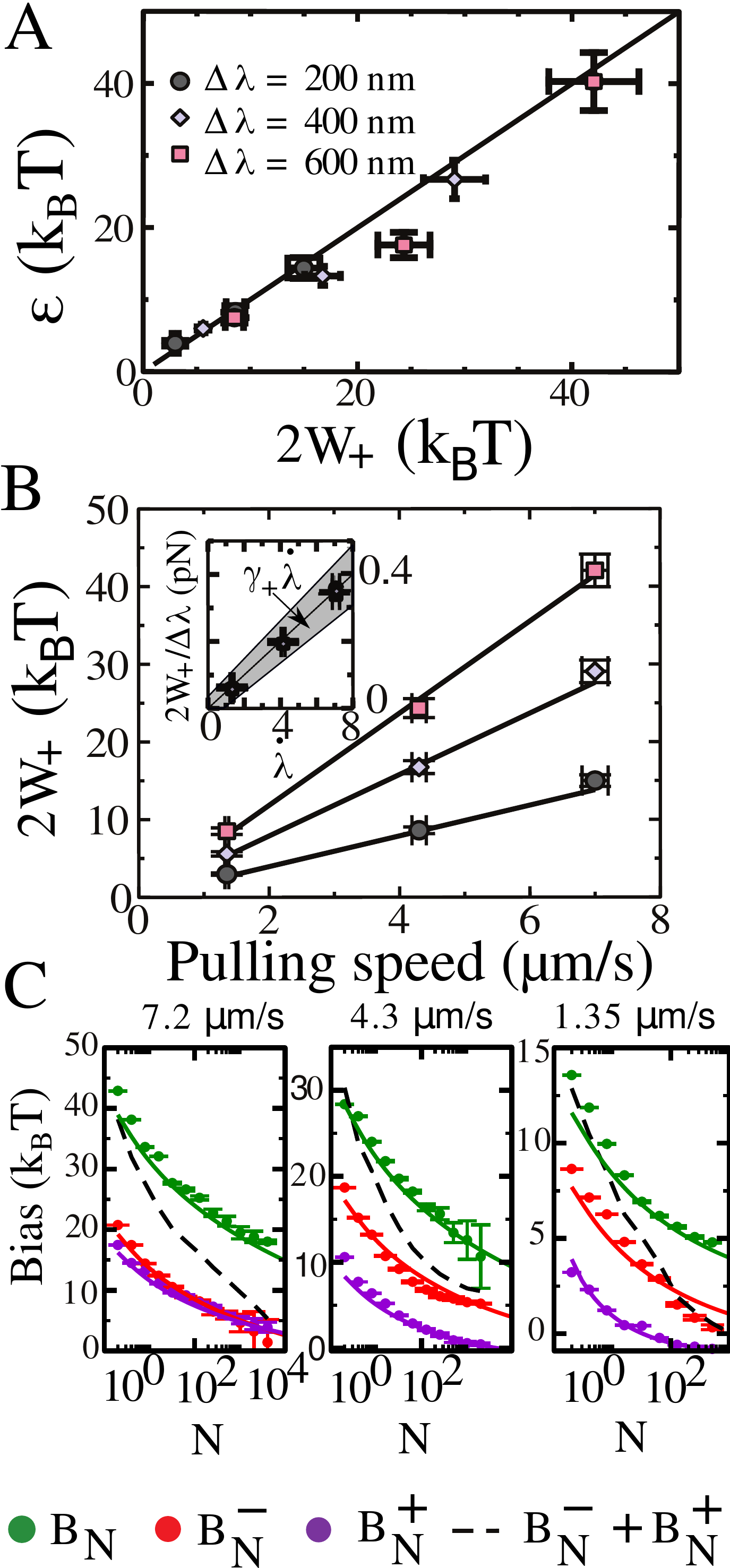}
 \caption{\label{Figure:seifert} }
 \end{figure}
%

{\bf Bias in unidirectional free energy estimators.} A) Error
($\mathcal{E}$) on free energy estimates Eq. \eqref{bias} committed by
using JE for $W'$.  Circles, diamonds and squares refer to excursions of
200 nm, 400 nm and 600 nm. Three different pulling speeds were
considered in each case. Note that in this case work is evaluated over
closed cycles ($\Delta G=0$) and the error is defined as:
$\mathcal{E}=-\beta^{-1}\log\langle e^{-\beta W'}\rangle$.  B) $\langle
W^+\rangle$ displays a bilinear dependence on pulling speed
$\dot\lambda$ (Main Figure) and $\Delta \lambda$ (Inset) as expected
from Eq. \eqref{defiw}. Continuous lines are linear fits to the
experimental results which contains a single fitting parameter. The
shaded area in the inset corresponds to the region within one standard
deviation from the expected value of $\langle W^+\rangle$ based on
equilibrium measurements of $\gamma_+$ (see Section 4,5 in the SI).  C)
Experimental bias measurements from the cycles shown in
Fig. \ref{Figure:opts}C. The plots show the bias
(Eqs. \eqref{fffd},\eqref{fffe}) as a function of the number of work
measurements, $N$. The three plots correspond to different pulling
speeds (7.2 $\mu$m/s, 4.3 $\mu$m/s, 1.35 $\mu$m/s). Interestingly
$B^-_N\ll B_N$ which guarantees faster convergence of free energy
estimates.  Moreover $B_N$ is also larger than the sum $B^-+B^+$ (dashed
line) i.e. the bias is superadditive (cf. Eq. \eqref{hiera}). The error
bars represent the statistical error on free energy determination, not
including systematic calibration errors in force and distance.
Continuous lines show the theoretical predictions from
Ref. \cite{prl.palassini.2011} for Gaussian work distributions. Note
that these are not fits but predictions which only use the mean
dissipation as input parameter.
%
%

\section*{Caption for Figure 4}

\begin{figure}[h]
  \centering
 \includegraphics[width=.7\textwidth]{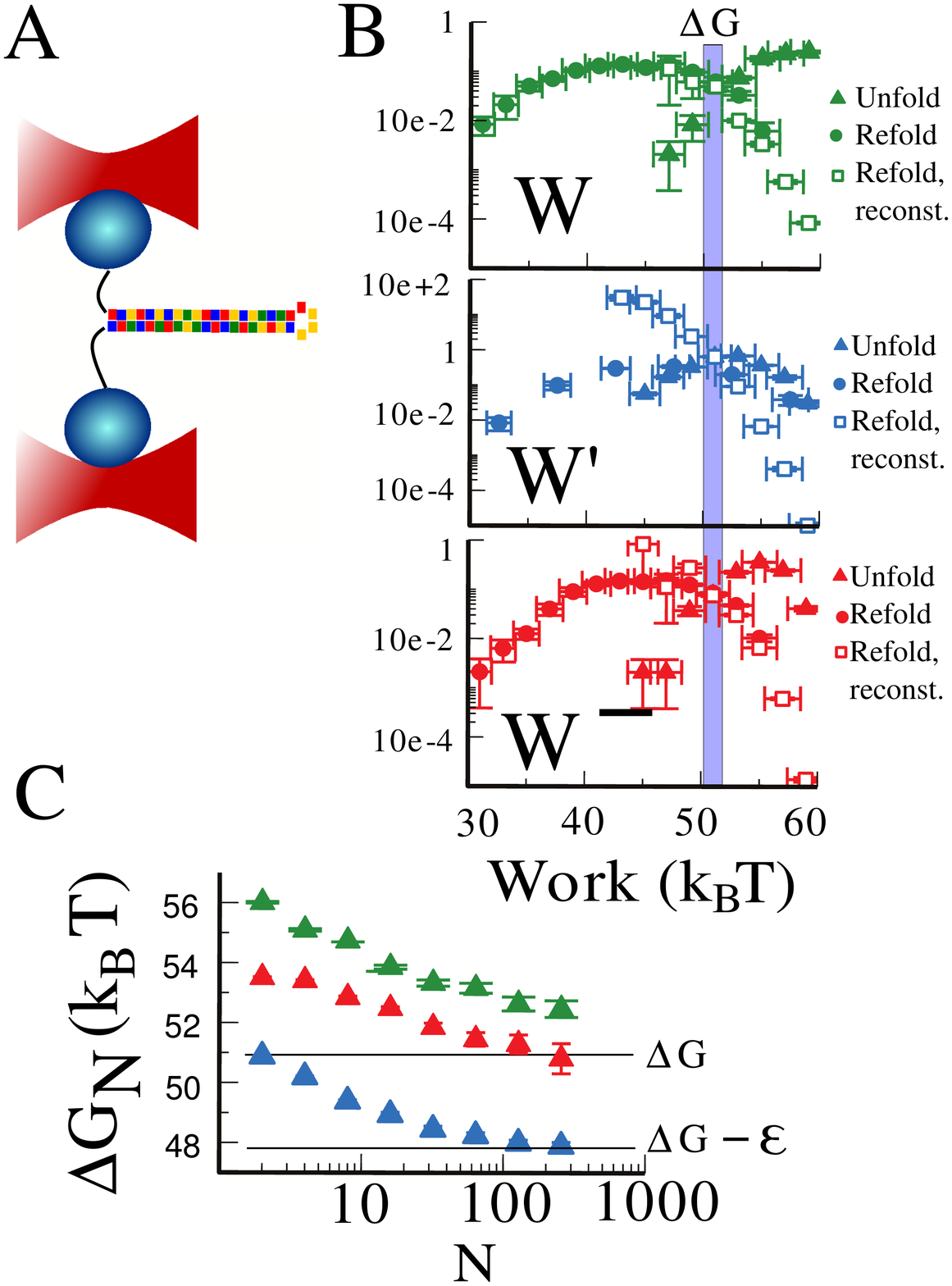}
 \caption{\label{Figure:sm}}
 \end{figure}

{\bf Work measurements on a DNA hairpin.} A)Scheme of the experimental setup (beads and hairpin not to scale). The hairpin is presented with color-coded sequence (A/T: yellow/green, G/C: red/blue).  B) Work measurements
upon unfolding and refolding according to three different work definitions: $W$ (upper panel), $W'$ (central panel) and $W^-$ (lower panel). In this experiment pulling speed was $0.96\pm0.02$ $\mu$m/s. Open symbols show an estimate of the refolding work distributions
reconstructed from the unfolding distribution via the CFR (Eq. \eqref{crozza}).  $W$ and $W^-$ fulfill the fluctuation theorem while $W'$ does not. Horizontal error bars represent the systematic error on work measurements, while vertical error bars denote statistical errors. The contribution of trap, handles and single stranded DNA have been removed as detailed in \cite{natphys.alemany.2012}.
C) Unidirectional estimates for the free energy from the unfolding work distribution. The optimal estimator based on $W^-$ (red) converges to the correct value $\Delta G_0=51$ $K_B$T as measured from bi-directional estimates. The estimator based on $W$ (green) shows a larger bias 
and overestimates the free energy by $\simeq1\,\, K_BT$. The estimator based on $W'$ (blue) converges to a wrong free energy difference ($\Delta G-\mathcal{E}$) which is $\simeq 3$ $K_BT$ below the correct value, against the second law. Note that the error committed by using $W$
is due to finite-size effects and decreases when more unfolding curves are measured. In contrast the error committed by using $W'$ remains finite for all sample sizes. 
The error bars represent the statistical error on free energy determination and do not include the systematic error due to force and distance calibrations.

\section*{Caption for Figure 5}

\begin{figure}[h]
 \centering
\includegraphics[width=0.99\textwidth]{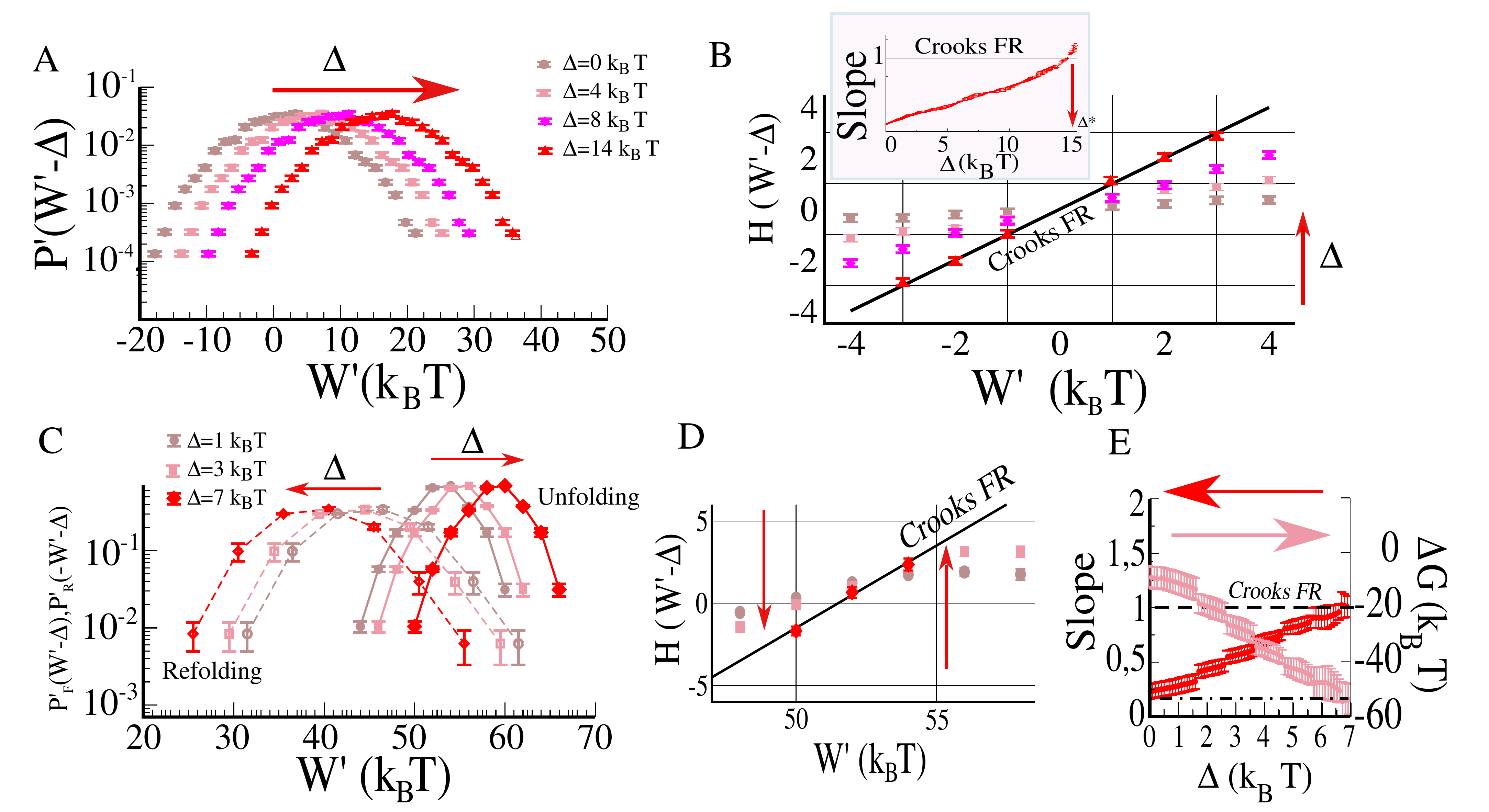}
\caption{\label{Figure5}}
 \end{figure}
%

{\bf Inference of $P(W)$ from partial
    work measurements in the symmetric case.} A) The distribution $P'(W')$  in the case of the dsDNA tether, for the work
  measured in the wrong trap, $W'$. The distribution does not fulfill the CFR. To recover the correct work distribution $P(W)$, 
  $W'$ is shifted by a constant amount $\Delta$. The shifted distribution is
  tested for the CFR by defining the function: $H(W')=\log
  (P'(W'-\Delta)/P'(-W'-\Delta))$.  The prediction by the CFR
  is $H(W')=\beta\left(W'-\Delta G\right)$ which can be tested
  by determining the $H$ function.  B) Evolution of  $H(W')$ as a function of $W'$ for different values of
  $\Delta$. The value $\Delta^*$ for which the slope of
  $H(W')$ is equal to one (work being measured in $K_BT$ units) determines the correct work
  distribution $P(W)$ ($\Delta\simeq 15$, inset). 
  C) In the case of bidirectional measurements both the forward and the reverse work
  distributions $P'_F(W'),P'_R(-W')$ are shifted by an amount $\Delta$
  in opposite directions. D) Evolution of the function
  $H=\log (P_F'(W'-\Delta)/P_R'(-W'-\Delta))$ as a function of
  $\Delta$. Again the CFR predicts $H$ should be linear in $W'$ 
  with slope one. E) Inference of the correct work distributions and
  $\Delta G$ measurement.  For each value of $\Delta$ a linear fit
  $A(\Delta)W+B(\Delta)$ to $H$ is
  performed. The value $\Delta^*$ for which $A(\Delta^*)=1$
  ($\Delta^*\simeq 7$ in the figure), is the shift needed in order to
  recover the full work distribution $P(W)$ from partial work measurements in
  the wrong trap. Moreover the CFR implies $B(\Delta^*)=-\Delta G$
  ($\Delta G\simeq 60K_BT$).

\section*{Caption for Figure 6}

\begin{figure}[h]
 \centering
 \includegraphics[width=0.7\textwidth]{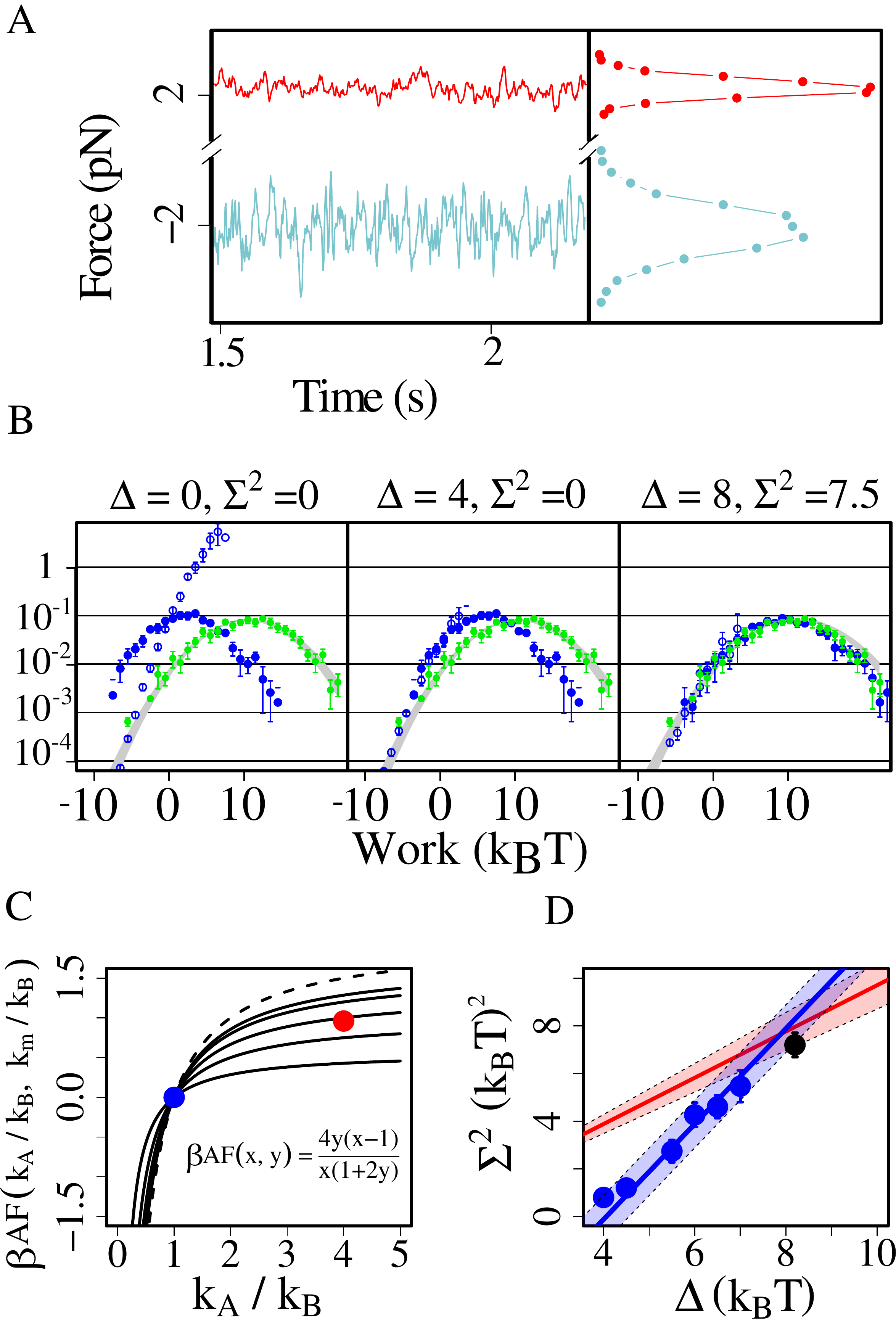}
 \caption{\label{Figure6}}
 \end{figure}

{\bf Inference of $P(W)$ from partial
    work measurements in the asymmetric case.} (A) Equilibrium force
distributions at $2$pN for the 3kb ds-DNA tether measured in an asymmetric dual-trap setup ($k_A = 0.012$
pN/nm,$k_B = 0.003$ pN/nm,$k_m = 0.0027$ pN/nm). (B) Convolution of
$P'(W')$ with different Gaussian distributions corresponding to
different pairs $\Delta,\Sigma$.We show $P'(W)$ (blue filled circles),
$P'(-W)\exp(\beta W)$ (blue empty circles), $P(W)$ (green filled circles).  
Among all different $P_{\Delta,\Sigma}$ only one matches the correct work distribution $P(W)$, i.e. only one reconstructed
distribution is physically correct (rightmost graph, with $ \Delta=7.5$
$K_B$T, $\Sigma^2 =8$ ($K_B$T$)^2$). In this
situation the inference cannot rest on the CFR alone,
and additional information is required to infer $P(W)$. (C) Asymmetry
factor (AF) as a function of $x = k_A/k_B$ for different values of $y =
k_m/k_B$. The blue (red) circles indicate the symmetric ($AF=0$) and
asymmetric ($AF\simeq 1$)
cases respectively. (D) The AF defined by $\Sigma^2=AF\times\Delta$ (red
line) and the CFR invariance
$\Delta=\Delta^*+\phi,\Sigma^2=\Sigma^{*2}+2\phi K_BT$ for any $\phi$
(blue line), do
select a narrow range of possible
pairs ($\Delta$,$ \Sigma$) at the intersection between the blue and red lines.
The intersection region is compatible with the parameters
($\Delta$ =8 $K_B$T, $ \Sigma^2$ =7.2 ($K_B$T)$^2$ ) describing the true
correct work distribution $P(W)$ (black point).

 \newpage
\clearpage


\section*{Supplementary material}

# Supporting information to: "Free energy inference from partial work measurements in small systems"

M. Ribezzi-Crivellari[1] and F. Ritort[1,2]

[1]Departament de Fisica Fonamental, Universitat de Barcelona
[2]Ciber-BBN de Bioingeneria, Biomateriales y Nanomedicina

## Contents

In this document we analyze in further detail several aspects of the manuscript regarding theory and experiments. Concerning the theory we study in detail a dynamical model describing the experiments reported in the main text. This model is then used to shed light on several aspects of the discussion in the main text. In section S1 we introduce the model in the symmetric case. In section S2 we prove that $x_+$ and $x_-$ are independent quantities in our model, which is a central result in our discussion. In section S3 we do the same for $W_+$ and $W_-$. These conclusions are supported by a statistical analysis of the experimental data. In sections S4 and S5 we comment on the definition and meaning of the hydrodynamic parameters of our model. One of these parameters is $\gamma_+$, which was used in estimating the missing contribution to the total dissipation when using the wrong work definition $W'$. We will also recall how to measure these parameters from equilibrium experiments and present measurements on different DNA tethers. In section S6 we will discuss free enerfy inference in asymmetric setups and in section S7 we present measurements at lower pulling speeds. These measurements show that correct work measurements is important even at low irreversibility conditions.

## S1 A MODEL FOR THE EXPERIMENTAL SETUP

We consider a model for the experimental dual-trap setup shown in Fig. S1. In this setup two focused laser beams form two optical traps, which allow us to manipulate a dumbbell formed by two trapped beads coupled through a molecular tether. The model is one dimensional: the dynamics takes place along the direction of the tether, $y$-axis in Fig. S1. Should the tether not be oriented in the plane perpendicular to the direction of the laser beams, the analysis is more involved as detailed in Ref. [1]. In this section we will assume the two traps are identical and harmonic, so that the total potential energy reads:

$$U_{\text{TOT}}(y_A, y_B) = \frac{k}{2} y_B^2 + \frac{k}{2}(y_A - \lambda)^2 + U_m(y_A - y_B). \tag{1}$$

A more general discussion, taking into account asymmetric setups, will be differed to section S6. Note that we choose the center of trap B (left trap) as the origin of our reference frame and that $\lambda$ is the position of trap A relative to trap B. A dynamical model for our system can be obtained using Langevin equations subject to a total potential as defined in Eq. (1):

$$\dot{\mathbf{y}} = -\bar{\mu} \nabla U_{\text{TOT}}(y_A, y_B, \lambda_t) + \boldsymbol{\eta}, \tag{2}$$

where $\mathbf{y} = (y_A, y_B)$, $\nabla = (\partial_{y_A}, \partial_{y_B})$, $\boldsymbol{\eta} = (\eta_A, \eta_B)$ is a vector formed by the noise terms affecting the two beads and $\bar{\mu}$ is the mobility tensor. We assume the noise to be $\delta$ correlated in time, and the Fluctuation-Dissipation Theorem to be fulfilled:

$$\langle \boldsymbol{\eta}_i(s) \boldsymbol{\eta}_j(t) \rangle = K_B T \mu_{i,j} \delta(t - s), \tag{3}$$

i.e. that the covariance of the noise is proportional to the mobility $\mu$. In the above expression $K_B$ is the Boltzmann constant and $T$ the absolute temperature. In the vector Equation (2) the mobility is a tensor or matrix. Its diagonal terms describe the friction affecting the motion of the beads, while the off-diagonal terms describe hydrodynamic interactions between the beads:

$$\bar{\mu} = \begin{pmatrix} \gamma_A^{-1}(y_A, y_B) & \Gamma_{AB}^{-1}(y_A, y_B) \\ \Gamma_{BA}^{-1}(y_A, y_B) & \gamma_B^{-1}(y_A, y_B) \end{pmatrix}.$$

The validity of Eq. (3) for hydrodynamic fluctuations in a pulling experiment is a standard assumption, and it is a necessary hypothesis in the derivation of FRs such as Jarzynski equality or the Crooks fluctuation relation. The analysis of this equation is greatly simplified by the following two features, which can be easily matched in our experimental setup:

- The setup is fully symmetric, meaning not only $k_A = k_B$ but also $r_A = r_B$ where $r_A, r_B$ are the bead radii.

- The hydrodynamic coefficients $\gamma, \Gamma$ between the beads do not significantly change in the force range of a pulling experiment. In section S4 we show that this condition is met for tethered molecules ranging from 20 nm to 1 $\mu$m in contour length to a 10% accuracy.

This guarantees that the mobility matrix is symmetric and constant:

$$\mu = \begin{pmatrix} \gamma^{-1} & \Gamma^{-1} \\ \Gamma^{-1} & \gamma^{-1} \end{pmatrix}.$$

Under these conditions, the transformation $x_+ = \frac{y_A + y_B}{2}, x_- = y_B - y_A$, decouples the potential into two separate terms,

$$U_{\text{TOT}}(x_-, x_+; \lambda) = U^-(x_-; \lambda) + U^+(x_+; \lambda).$$

Moreover the symmetry requirements cited above have the following two important consequences:

- It diagonalizes the mobility tensor $\mu$ in the form:

$$\mu = \begin{pmatrix} \gamma_+^{-1} & 0 \\ 0 & \gamma_-^{-1} \end{pmatrix}.$$

- It diagonalizes the noise covariance, i.e. it gives rise to two statistically independent noise contributions, $\eta_+$ and $\eta_-$ which separately affect the new coordinates $x_+$ and $x_-$:

$$\langle \eta_\pm(s)\eta_\pm(t) \rangle = K_B T \gamma_\pm \delta(t-s), \qquad \langle \eta_+(s)\eta_-(t) \rangle = 0. \tag{4}$$

In the main text we have shown how the partition functions for $x_-$ and $x_+$ decouple. From previous equations one can easily see that the full right-left symmetry of the setup guarantees that also the equations of motion for $x^+$ and $x^-$ decouple completely:

$$\gamma_+ \dot{x}_t^+ = -\partial_{x^+} U^+(x_t^+, \lambda_t) + \eta_t^+ \tag{5}$$

$$\gamma_- \dot{x}_t^- = -\partial_{x^-} U^-(x_t^-, \lambda_t) + \eta_t^-. \tag{6}$$

## S2 THE DIFFERENTIAL $X_-$ AND CENTER-OF-MASS $X_+$ COORDINATES ARE STATISTICALLY INDEPENDENT

The factorization of the partition function (Eq.8 in the main text) plays an important role in the derivation of our results. For symmetric systems such factorization is directly related to the statistical independence of the center of mass and differential coordinates. Here we experimentally check this factorization by showing how we can generally describe the state of our system using a decomposition of the measured forces into two statistically independent coordinates. For linear systems such decomposition is always possible whereas for general non-linear systems such decomposition is only possible in the symmetric case. Imagine that the equilibrium distribution of our system is dictated by a Hamiltonian depending on two coordinates: $U(y_A, y_B)$, which takes the form $U_+(x_+) + U_-(x_-)$ for a general transformation, $x_- = x_-(y_A, y_B)$ and $x_+ = x_+(y_A, y_B)$. In this situation, $x_+$ and $x_-$ are statistically independent:

$$P(x_+|x_-) = P(x_+) \propto \exp\left(-\beta U_+(x_+)\right), \tag{7}$$

and in particular have zero covariance:

$$\langle x_+ x_- \rangle = \langle x_+ \rangle \langle x_- \rangle. \tag{8}$$

A further consequence of this decomposition of the Hamiltonian is the factorization of the partition function discussed in the main text. For simplicity we will be considering forces instead of distances, as our instrument directly measures forces. The independence of the coordinates $x_+$ and $x_-$ directly translates into the independence of the corresponding forces:

$$f_+ = -\partial U_+(x_+) \qquad f_- = -\partial U_-(x_-), \tag{9}$$

indeed:

$$\langle f_+ f_- \rangle = \int dx_+ dx_- f_+ f_- e^{-\beta U} = \int dx_+ f_+ e^{-\beta U_+} \int dx_- f_- e^{-\beta U_-} = \langle f_+ \rangle \langle f_- \rangle. \tag{10}$$

We shall use the covariance as a measure of linear dependence and a Pearson's $\chi^2$ test as a measure of dependence in general. The $\chi^2$ test is described below in section S3 of this document. The transformations we use to uncouple the two degrees of freedom are rotations of the form:

$$f_+^\phi = \cos(\phi) f_A + \sin(\phi) f_B \tag{11}$$

$$f_-^\phi = -\sin(\phi) f_A + \cos(\phi) f_B \tag{12}$$

We present measurements of equilibrium force fluctuations obtained using different tethers. In Figure S2 we describe measurements performed on 1kb dsDNA tethers stretched at 2 pN. Experiments were performed using PBS buffer (pH 7.5), 1M NaCl at 25 Celsius degrees just as in the pulling experiments described in the main text. The force of 2pN was chosen to fall in the range explored in the pulling experiments. The first set of measurements was performed using symmetric traps $k_A \simeq k_B \simeq 0.02$ pN/nm. The second set of experiments was performed with asymmetric traps, $k_A = 0.012$ pN/nm, $k_B = 0.003$ pN/nm. In both cases it is possible to find a coordinate system such that the two coordinates are independent and in both cases this coordinate system is obtained by a rotation of the vector $(f_A, f_B)$ (Fig. S2A,C). The difference is that in the symmetric case the new variables correspond to the center-of-mass and differential coordinate as discussed in the main text (Fig S2B), while they have a different definition (i.e. they are not generated by a $\pi/4$ rotation) in the asymmetric case (Fig. S2D). In Figure S3 we report similar measurements performed on a DNA hairpin (the sequence is the same reported in the main text) in a symmetric setup. Equilibrium traces were acquired at different forces. From top to bottom we can see the hairpin is 1) completely folded, 2) preferentially folded, 3) preferentially unfolded, 4) completely unfolded. Force fluctuations are clearly non-Gaussian, a characteristic of non-linear two-state systems. Again, for the symmetric setup, our data show that the center of mass $(x_+)$ and differential coordinate $(x_-)$ are independent. We note that in the more general non-linear case and for asymmetric setups no independent coordinates can be defined.

## S3 THE DIFFERENTIAL WORK $W_-$ AND $W_+$ ARE STATISTICALLY INDEPENDENT

An immediate consequence of the decoupling of the equations of motion is the independence of $W_+$ and $W_-$:

$$W_\pm = \int \partial_\lambda U_\pm(x_\pm, \lambda) \dot\lambda dt.$$

Now, since $W^+$ does only depend on $x^+$ and $W^-$ does only depend on $x^-$ we can conclude that the two quantities are independent. Of course this results holds under specific assumptions and is, in the end, just a property of the model. From the experimental point of view one can test the independence of $W^+$ and $W^-$ by statistical analysis, studying their linear (Pearson) correlations or using a Pearson's $\chi^2$ test for dependence. The first method is sensitive to linear correlation between two observables, while the second, more stringent test, is sensitive to generic correlations. The $\chi^2$ test goes as follows: to test the independence of two random variables $(A, B)$ from a finite number of measurements we first define the empirical marginals, i.e. separate histograms of the two variables:

$$e_j^a = \sum_i^N \delta(A_i \in [a_j, a_{j+1})) / N \tag{13}$$

$$e_j^b = \sum_i^N \delta(B_i \in [b_j, b_{j+1})) / N, \tag{14}$$

here $\delta(x \in \mathcal{C})$ is 1 if $x$ belongs to the set $\mathcal{C}$ and is 0 otherwise. We then compare the actual two-dimensional histogram, $e_{ij}$ to the product of the marginals:

$$\chi^2 = \sum_{k,l} \frac{\left(e_{kl} - e_k^a e_l^b\right)^2}{e_k^a e_l^b},\tag{15}$$

with $e_{jk} = \sum_i \delta(A_i \in [a_j, a_{j+i}))\delta(B_i \in [b_k, b_{k+i}))/N$. The value obtained through Eq. (15) is then compared to the $\chi^2$ distribution. The two variables are considered independent if the $\chi^2$ value satisfies 5% significance test. The results of the two statistical tests is reported in Table S1 and S2. These tables compare the result of the tests performed on the pair $W, W'$ or on the pair $W_+, W_-$, for the forward, the reverse and the cyclic protocols. In each case three different pulling speeds are considered. The two different tables refer to two different dumbbells. In most cases the pair $W, W'$ shows higher covariance than the pair $W_+, W_-$. Moreover the $\chi^2$ test detects dependence between $W, W'$ but not between $W_+, W_-$ (in the $\chi^2$ column 1 denotes a positive test for dependence and 0 a negative test).

## S4 DIRECT MESUREMENT OF THE HYDRODYNAMIC PARAMETERS

In the main text we have already shown how, in a symmetric setup, by changing to the coordinate system defined by the center of mass and differential coordinates $x_+, x_-$, the potential decouples into two terms:

$$\begin{aligned} U_{\text{TOT}}\left(x_+, x_-\right) = {} & k(x_+ - \frac{\lambda}{2})^2 \\ & + \frac{k}{4}(x_- - \lambda)^2 + U_m(x_-). \end{aligned}\tag{16}$$

When considering equilibrium fluctuations at constant trap–to–trap distance $\lambda$, a linear approximation of $U_m$ in Eq. (16) may be used. The energy has a minimum at $x_+ = \lambda$, $x_- = x_-^0$. The value of $x_-^0$ is defined by:

$$\partial_{x_-} U_{\text{TOT}}\bigg|_{x_-^0} = 0.\tag{17}$$

Here $x_-^0$ is the (mechanical) equilibrium distance between the centers of the beads. The position of the minimum and the derivative of the potential at the minimum depend on the trap–to–trap distance, so that $x_-^0$ is a function of $\lambda$. In this approximation the equations of motion in Eq. (5) become:

$$\gamma_+ \dot{x}_+ = -2k\left(x_+ - \frac{\lambda}{2}\right) + \eta_+,\tag{18}$$

$$\gamma_- \dot{x}_- = -\left(\frac{k}{2} + k_m\right)\left(x_- - x_-^0\right) + \eta_-.\tag{19}$$

In particular $\gamma_+$ is the friction affecting the center of mass of the dumbbell, while $\gamma_-$ affects the dynamics of the differential coordinate. After Bachelor [3] we set:

$$\gamma_+^{-1} = \left(\gamma^{-1} + \Gamma^{-1}\right)/2, \qquad \gamma_-^{-1} = 2\left(\gamma^{-1} - \Gamma^{-1}\right).\tag{20}$$

In brief, $\gamma$ is the hydrodynamic friction coefficient for a single bead, while $\Gamma$ is the intensity of hydrodynamic interactions between the two beads. It is important to bear in mind that in principle $\gamma$ and $\Gamma$ depend on the differential coordinate $x_-$. This dependence is negligible in pulling experiments which induce a small change in $x_-$, but it becomes clear and measurable if $x_-$ is changed to a larger extent, e.g. using molecules of different contour length. The equilibrium probabilities generated by (18),(19) are given by Boltzmann's distribution:

$$Q_{eq}(x_+) = \frac{1}{Z_+} \exp\left(-\beta k(x_+ - \frac{\lambda}{2})^2\right),\tag{21}$$

$$P_{eq}(x_-) = \frac{1}{Z_-} \exp\left(-\beta \frac{k + 2k_m}{4}\left(x_- - x_-^0\right)^2\right),\tag{22}$$

with $Z_+, Z_-$ normalization factors. The variance of equilibrium fluctuations in $x_+$ and $x_-$ is connected to the elastic properties of traps and tether by:

$$\sigma_+^2 = \frac{K_B T}{2k}, \qquad \sigma_-^2 = \frac{2K_B T}{k + 2k_m}.$$
(23)

Information about the hydrodynamic interactions can be obtained from the time-dependent correlation functions of $x_+$ and $x_-$:

$$C_+(t) = \quad \langle (x_+(t) - \tfrac{\lambda}{2})(x_+(0) - \tfrac{\lambda}{2}) \rangle$$
(24)

$$C_-(t) = \quad \langle (x_-(t) - x_-^0)(x_-(0) - x_-^0) \rangle,$$
(25)

which characterizes the decay of fluctuations and allows to distinguish the presence of different contributions to the total variance. The computation of the correlation functions yields:

$$\beta C_+(t) = \frac{e^{-k\gamma_+^{-1} t}}{2k},$$
(26)

$$\beta C_-(t) = \frac{2 e^{-(k+2k_m)\gamma_-^{-1} t}}{k + 2k_m},$$
(27)

where $\beta = (K_B T)^{-1}$. From Eqs. (24),(25) the correlation function at time 0 equals the variances given in Eq. (23):

$$C_+(0) = \sigma_+^2, \qquad C_-(0) = \sigma_+^2.$$
(28)

The hydrodynamic parameters can be retrieved from the correlation functions, as the stiffness, $k, k_m$, are known from Eq. (23) :

$$\frac{1}{\gamma} + \frac{1}{\Gamma} = \left. -\frac{1}{2k} \frac{d}{dt} \log(C_+) \right|_{t=0}.$$
(29)

$$\frac{1}{\gamma} - \frac{1}{\Gamma} = \left. -\frac{2}{k + 2k_m} \frac{d}{dt} \log(C_-) \right|_{t=0}.$$
(30)

where we used Equations (20),(26) and (27). As discussed in the main text, $\gamma_+$ is needed to correct the error committed by using the force measured in the trap at rest in the Jarzynski Equality (Eq. 17 and 18 in the main text).

## S5 A CLOSER LOOK TO HYDRODYNAMIC INTERACTIONS

In the previous section we have shown how the analysis of thermal fluctuations in a dual trap setup allows the measurement of the two scalar parameters entering in the hydrodynamics: $\gamma^{-1}$ and $\Gamma^{-1}$. The sum of the two parameters can be obtained through Eq. (29) from the analysis of the time correlation function for $x_+$, Eq. (26) . The difference is instead obtained by a similar analysis on the decay rate of $C_-(t)$. Up to now we have neglected the dependence of the hydrodynamic parameters on the relative distance between the two beads. This is justified provided this relative distance does not change too much in the pulling experiments. Conversely changing the contour length of the molecule by an order of magnitude leads to a larger and detectable change in the hydrodynamic parameters. This offers the possiblity of testing our measurements against the theoretical prediction obtained using Stokes equation and expressed as a power series in the reduced distance $\rho = \langle x_- \rangle / r_b$ with $r_b$ the radius of the beads and $\langle x_- \rangle$ the mean of the differential coordinate. (Note that $\langle x_- \rangle > 2r_b$ and thus $\rho > 2$). The two quantities can be computed to arbitrary precision, taking the first two terms of the series expansion reported in [4]:

$$\gamma^{-1} \simeq \frac{1}{6\pi\eta r_b} \left( 1 - \frac{15}{4\rho^4} + O(\rho^{-6}) \right)$$
(31)

$$\Gamma^{-1} \simeq \frac{1}{6\pi\eta r_b} \left( \frac{3}{2\rho} - \frac{1}{\rho^3} + O(\rho^{-7}) \right).$$
(32)

We measured fluctuations using ds-DNA tethers of 4 different contour lengths: 8 $\mu$m (24 kbp), 1 $\mu$m (3 kbp), 300 nm (1.2 kbp) and 20 nm (58 bp). For each tether fluctuations were measured at different forces. The different length of the tethers used in the experiments is such that $\rho$ assumes values from above 6 down to 2.03, near the minimum value 2 taken when the beads are in contact: hydrodynamic interactions are monitored from the far–field regime ($\rho \gg 2$) to the lubrication limit ($\rho \simeq 2$). The experimental data for $\gamma$ and $\Gamma$ can be confronted with predictions Eq. (31),(32) without free parameters as the buffer viscosity and the bead radius are known.

This comparison is shown in Figure S4, the upper plot compares the measured values for $\gamma^{-1}$ and $\Gamma^{-1}$ with Eq. (31),(32) (solid lines). For every tether we assumed a single value of $\rho$ and averaged values obtained at different forces. On the force range relevant for pulling experiments, changes in $\gamma^{-1}$ and $\Gamma^{-1}$ where below 10%. Theory and data do agree, within the error bars for the longer tethers. The data for the shortest tether show a deviation from expression (31) and probably more terms of the expansion would be needed. Nevertheless in this case the parameters are very close to the contact value, $\rho = 2$, $\gamma^{-1}(2)/\gamma^{-1}(\infty) = \Gamma^{-1}(2)/\gamma^{-1}(\infty) = 0.775$. The precision of these measurements is limited, in our case, by the error with which the radius of the bead is known. One can reduce the dependence of the measured value on the size of the bead by using the ratio between the two parameters, which does not depend on $r_b$ if not through the definition of $\rho$. The results are shown in the lower plot where the measured data are compared to those obtained from untethered beads. This allows us to conclude that, at least in our conditions and within our experimental resolution, the friction and/or hydrodynamic effect due to the presence of a polymer between the beads is not distinguishable. In particular, knowing the value of $\gamma_+$ allows an estimation of the systematic error on unidirectional free energy estimates from single molecule pulling experiments based on wrong work definitions, as detailed in the main text.

## S6 FREE ENERGY INFERENCE IN ASYMMETRIC SETUPS

In the case of asymmetric setups $P(W)$ and $P(W')$ do not have such a simple relationship as in the symmetric case. In this case a successful inference cannot be based on the FR alone: some additional, system-specific, information must be provided. We shall focus on asymmetric systems in the Gaussian approximation and use pulling experiments on dsDNA in an asymmetric setup to test our predictions. Is it still possible to infer the full dissipation from partial work measurements? The answer is positive if we are given some equilibrium information on the system. In this case it is enough to know the trap and molecular stiffnesses $k_B, k_A, k_m$ i.e. equilibrium properties of the system. But why does direct inference fail in the asymmetric case? When discussing symmetric systems we have shown that $P(W)$ and $P(W')$ are related by a simple shift:

$$P(W) = P'(W - \Delta).\tag{33}$$

Only the mean of the probability distribution had to be changed, as the variance of the two distributions is the same. In that case imposing the validity of the CFR for $P(W)$ yields the unique value of $\Delta$ to be used in the reconstruction. This is not true anymore in asymmetric systems. Here both the mean and the variance of the work distribution must be changed, which can be achieved by convolution:

$$P(W)_{\Delta,\Sigma} = P' \star \mathcal{N}(\Delta, \Sigma),\tag{34}$$

where $\star$ denotes the convolution operator and $\mathcal{N}(\Delta, \Sigma)$ is a normal distribution with mean $\Delta$ and standard deviation $\Sigma$. Starting from any distribution $P'(W')$ there are infinitely many choices of $\Delta$ and $\Sigma$ which yield a $P(W)_{\Delta,\Sigma}$ satisfying the fluctuations theorem. Indeed, let us suppose the pair $\Delta^*, \Sigma^*$ is such that:

$$P_{\Delta^*,\Sigma^*}(W) = P' \star \mathcal{N}(\Delta^*, \Sigma^*)\tag{35}$$

satisfies the CFR. Then it is easy to check that $P_{\Delta^*+\phi, \sqrt{(\Sigma^{*2}+2\phi/\beta)}}$ will satisfy the FT for any $\phi$ ($\beta = 1/K_B T$ as always). In Figure S5 we show the effect of the convolution of $P'(W')$ with different Gaussian distributions. In the rightmost column of that plot we highlight three different pairs for which the convolved distribution satisfies the CFR. In this situation the inference cannot rest on the CFR alone, and additional information about the system is needed. In the general scheme of a Gaussian approximation,

inference of $P(W)$ is still possible. In the coming subsections we will show that $P(W)$ and $P'(W')$ are related by an Asymmetry Factor (AF) given by:

$$AF(k_A, k_m, k_B) = \frac{\sigma_W^2 - \sigma_{W'}^2}{\langle W \rangle - \langle W' \rangle} = \frac{1}{\beta} \frac{4k_m(k_A - k_B)}{k_A(k_B + 2k_m)}, \tag{36}$$

where $\beta = 1/K_B T$. The $AF$ is an equilibrium quantity: it only depends on the stiffnesses of traps and tether. Knowing the $AF$ allows us to select the unique pair $(\Delta, \Sigma)$ such that $AF = \Sigma^2/\Delta$ and that $P_{\Delta, \Sigma}(W)$ satisfies the fluctuation symmetry. These $(\Delta, \Sigma)$ allow the reconstruction of $P(W)$ and thus to measure free energy differences or even dynamical quantities like $\gamma_+$. The behavior of the $AF$ factor as a function of $x = k_B/k_A$ and $y = k_m/k_A$ is shown in the left panel of Figure S6.

## S6.1 PULLING EXPERIMENTS IN LINEAR SYSTEMS

Free energy inference in asymmetric setups is not as straightforward as it is in symmetric setups. In order to recover the full dissipation from partial work measurements we must complement these non-equilibrium measurements with some equilibrium information about the system. In this section we will consider a Gaussian approximation for asymmetric setups that, as shown in Figure 6 in the main text, agrees pretty well with the experimental results. The Gaussian case will be modelled by linear asymmetric systems, a class of statistical models for which inference is possible without symmetry restrictions. This amounts to choosing $U_m(x) = \frac{1}{2} k_m x^2$ in Eq. 1 so that the total potential reads:

$$U(y_A, y_B, \lambda) = \frac{k_m}{2}(y_B - y_A)^2 + \frac{k_B}{2} y_B^2 + \frac{k_A}{2} (\lambda - y_A)^2. \tag{37}$$

In order to simplify the following discussion we will now switch to a vector notation. From now on we shall denote with $\mathbf{X}^T(t)$ the vector containing the positions of the traps (remind that B is the reference trap at rest with respect to water while trap A moves):

$$\mathbf{X}^T(t) = (\lambda_t, 0), \tag{38}$$

and with $\mathbf{y}$ the vector of bead positions: $\mathbf{y} = (y_A, y_B)$. In this vector notation the potential can be written in its normal form as:

$$U(\mathbf{y}, \mathbf{X}^T(t)) = \frac{1}{2}(\mathbf{y} - \mathbf{y}^0) \cdot \bar{\mathsf{k}} (\mathbf{y} - \mathbf{y}^0) + \frac{1}{2} k_{\text{eff}} \lambda^2, \tag{39}$$

where $\bar{\mathsf{k}}$ is the stiffness tensor

$$\bar{\mathsf{k}} = \begin{pmatrix} k_A + k_m & -k_m \\ -k_m & k_B + k_m \end{pmatrix}, \tag{40}$$

$k_{\text{eff}}$ is the effective stiffness of the dumbell as a whole:

$$k_{\text{eff}} = \frac{k_A k_B k_m}{\det(\bar{\mathsf{k}})}, \tag{41}$$

and $\mathbf{y}^0$ is the vector whose component are the equilibrium positions of the beads:

$$\mathbf{y}^0 = \left( \lambda - \frac{k_B k_m}{\det(\bar{\mathsf{k}})}(\lambda), \frac{k_A k_m}{\det(\bar{\mathsf{k}})}(\lambda) \right). \tag{42}$$

During a pulling experiment the trap to trap distance will be changed at a constant speed, $v$:

$$\lambda_t = \lambda_0 + vt. \tag{43}$$

As a consequence the equilibrium positions of the beads will change in time:

$$\mathbf{y}^0(t) = \left( \lambda_0 + vt - \frac{k_B k_m}{\det(\bar{\mathsf{k}})}(\lambda_0 + vt), \frac{k_A k_m}{\det(\bar{\mathsf{k}})}(\lambda_0 + vt) \right). \tag{44}$$

During a pulling experiment the free energy of the system changes in time. FRs can be used to determine the free energy change in the system from the distribution of the work performed on the system. As discussed in the main text, the correct definition of the work in such an experiment depends on the way in which the traps are moved. Work is identified as the time derivative of the total potential $U$:

$$W = -v \int_0^t k_A \left( y_A(s) - X_A^T(s) \right) ds \tag{45}$$

Any different definition of work does not yield reliable free energy differences. An alternative work definition can be based on the force measured in the trap at rest (B):

$$W' = v \int_0^t k_B \left( y_B(s) - X_B^T(s) \right) ds, \tag{46}$$

a choice made by many experimentalists. Is it possible to infer the distribution of $W$ from the distribution of $W'$ and thus to reliably measure free energy differences? The answer to this is worked out in detail below.

## S6.2 W AND W' AS GAUSSIAN RANDOM VARIABLES

In order to answer the questions posed in the previous section we need to study the distributions of $W$ and $W'$. In this linear model $y_A$ and $y_B$ are Gaussian random variables, so that $W$ and $W'$, being linear in $y_A$ and $y_B$, are themselves Gaussian random variables too. The distribution of a Gaussian random variable is determined by its mean and variance. The computation of the mean and variance of the work requires the solution of a system of linear differential equations that can be obtained form the Markov Generator for the joint process $(\mathbf{y}, \mathbf{W})$ or equivalently from the corresponding Fokker-Planck equation. Here by $\mathbf{y}$ we denote the usual vector containing the positions of the beads while $\mathbf{W} = (v \int_0^t f_A(s)ds, v \int_0^t f_B(s)ds)$. We use the vector $\mathbf{W}$ from which the random variables $W, W'$ are obtained as:

$$W = \mathbf{W}_A, \qquad W' = -\mathbf{W}_B. \tag{47}$$

Just as we did when discussing the measurement of hydrodynamic parameters we will use Langevin equations to describe the dynamics of our system:

$$\dot{\mathbf{y}} = -\bar{\mu}\bar{\mathbf{k}}(\mathbf{y} - \mathbf{y}^0) + \eta, \tag{48}$$

where $\bar{\mu}$ is the mobility tensor, describing both friction and hydrodynamic interactions:

$$\bar{\mu} = \begin{pmatrix} \gamma^{-1} & \Gamma^{-1} \\ \Gamma^{-1} & \gamma^{-1} \end{pmatrix}, \tag{49}$$

and $\eta$ is a white noise compatible with the fluctuation-dissipation theorem, i.e. $\langle \eta_t \eta_s \rangle = 2\beta^{-1}\bar{\mu}\delta(t-s)$. In this setting the Markov generator $L$ for the process $(\mathbf{y}, \mathbf{W})$ is given by:

$$\begin{aligned} Lf(\mathbf{y}, \mathbf{W}) = \\ = \left( -\bar{\mu}\bar{\mathbf{k}}(\mathbf{y} - \mathbf{y}^0) \cdot \nabla_y + \beta^{-1}\bar{\mu} \cdot \nabla_y \nabla_y - v\bar{\mathbf{k}}_D(\mathbf{y} - \mathbf{X}^T) \cdot \nabla_w \right) f(\mathbf{y}, \mathbf{W}), \end{aligned} \tag{50}$$

where

$$\bar{\mathbf{k}}_D = \begin{pmatrix} k_A & 0 \\ 0 & k_B \end{pmatrix}. \tag{51}$$

We recall that the Markov generator is the infinitesimal evolution operator in the sense that:

$$\frac{\partial}{\partial t} \langle f(\mathbf{y}(t), \mathbf{W}(t)) \rangle_y = \langle Lf(\mathbf{y}(t), \mathbf{W}(t)) \rangle_y, \tag{52}$$

where $\langle \rangle_y$ denotes average conditioned to initial condition $(y, 0)$. Applying the Markov Generator to $\mathbf{y}, (\mathbf{y} - \langle \mathbf{y} \rangle_y) \otimes (\mathbf{y} - \langle \mathbf{y} \rangle_y), \mathbf{W}, (\mathbf{W} - \langle \mathbf{W} \rangle_y) \otimes (\mathbf{W} - \langle \mathbf{W} \rangle_y)$ [1] and then taking the average, we obtain the

---

[1] $\mathbf{i} \otimes \mathbf{j}$ denotes the tensor $\bar{\mathbf{J}}$ such that $\bar{\mathbf{J}}\mathbf{k} = (\mathbf{i} \cdot \mathbf{k})\mathbf{j}$.

equations of motion for $\langle\mathbf{y}\rangle_y$, $\bar{\sigma}_y^2 = \langle(\mathbf{y} - \langle\mathbf{y}\rangle_y) \otimes (\mathbf{y} - \langle\mathbf{y}\rangle_y)\rangle_y$, $\langle\mathbf{W}\rangle_y$, $\bar{\mathsf{C}} = \langle(\mathbf{W} - \langle\mathbf{W}\rangle_y) \otimes (\mathbf{y} - \langle\mathbf{y}\rangle_y)\rangle_y$ and $\bar{\sigma}_W^2 = \langle(\mathbf{W} - \langle\mathbf{W}\rangle_y) \otimes (\mathbf{W} - \langle\mathbf{W}\rangle_y)\rangle_y$.

$$\frac{d}{dt}\langle\mathbf{y}\rangle_y = -\bar{\mu}\bar{\mathsf{k}}(\langle\mathbf{y}\rangle_y - \mathbf{y}^0(t)) \tag{53}$$

$$\frac{d}{dt}\bar{\sigma}_y^2 = -2\mathrm{Sym}\left(\bar{\mu}\bar{\mathsf{k}}\bar{\sigma}_y^2 + \frac{\bar{\mu}}{\beta}\right) \tag{54}$$

$$\frac{d}{dt}\langle\mathbf{W}\rangle_y = -\bar{\mathsf{k}}_D(\langle\mathbf{y}\rangle_y - \mathbf{X}^T(t)) \tag{55}$$

$$\frac{d}{dt}\bar{\mathsf{C}} = -\bar{\mu}\bar{\mathsf{k}}\bar{\mathsf{C}} - \bar{\sigma}_y^2\bar{\mathsf{k}}_D \tag{56}$$

$$\frac{d}{dt}\bar{\sigma}_W^2 = -2\bar{\mathsf{k}}_D\bar{\mathsf{C}}, \tag{57}$$

where $\mathrm{Sym}(\cdot)$ in (54) is the operator which gives the symmetric part of its argument. These equations are explicitely derived in subsection S6.5 and solved in subsection S6.6, while we shall now comment on the result.

The calculations reported in the next sections show that, after neglecting transients:

$$\langle\mathbf{W}(t)\rangle = -v\int_0^t \bar{\mathsf{k}}_D\left(\mathbf{y}^0(s) - \mathbf{X}^T(s)\right)ds + \bar{\mathsf{k}}_D\bar{\mathsf{k}}^{-1}\bar{\mu}^{-1}\bar{\mathsf{k}}^{-1}\bar{\mathsf{k}}_D\dot{\mathbf{X}}^T vt, \tag{58}$$

and

$$\bar{\sigma}_W^2 = 2\frac{\bar{\mathsf{k}}_D\bar{\mathsf{k}}^{-1}\bar{\mu}^{-1}\bar{\mathsf{k}}^{-1}\bar{\mathsf{k}}_D}{\beta}v^2 t \tag{59}$$

The first term on the right hand side of (58) is the integral:

$$v\bar{\mathsf{k}}_D\left(\int_0^t -\left(\mathbf{y}^0(s) - \mathbf{X}^T(s)\right)ds\right), \tag{60}$$

the difference $\bar{\mathsf{k}}_D(\mathbf{y}^0(s) - \mathbf{X}^T(s))$ gives the equilibrium force in each trap at time $s$, this is just the force that would be obtained at the corresponding trap-to-trap distance $\lambda$ if the pulling were carried out reversibly. The two components of the vector $\bar{\mathsf{k}}_D\left(\int_0^t \left(\mathbf{y}^0(s) - \mathbf{X}^T(s)\right)ds\right)$ are of course equal in size but different in sign:

$$v\bar{\mathsf{k}}_D\left(\int_0^t -\left(\mathbf{y}^0(s) - \mathbf{X}^T(s)\right)ds\right) = \left(v\int_0^t f_{\mathrm{rev}}(s)ds, -v\int_0^t f_{\mathrm{rev}}(s)ds\right). \tag{61}$$

The above expression is then just the reversible work:

$$W_{\mathrm{rev}} = k_{\mathrm{eff}}\left(\lambda_0 vt + \frac{1}{2}v^2 t^2\right). \tag{62}$$

where we used Eqs.(39) and (43). The second term in (58) is the dominant non-equilibrium contribution (by the way, this is the only term which is linear in time), which arises from the finite pulling speed of the protocol. A key element in this dissipation term is:

$$\bar{\Omega} = \bar{\mathsf{k}}_D\bar{\mathsf{k}}^{-1}\bar{\mu}^{-1}\bar{\mathsf{k}}^{-1}\bar{\mathsf{k}}_D = \bar{\mathsf{p}}^T\bar{\mu}^{-1}\bar{\mathsf{p}}. \tag{63}$$

To give some insight on the physical meaning of this quantity we note that $\bar{\mathsf{p}}$ transforms the trap position vector $\mathbf{X}^T$ into the equilibrium position vector $\mathbf{y}^0$:

$$\mathbf{y}^0 = \bar{\mathsf{p}}\mathbf{X}^T. \tag{64}$$

This dissipation term stems from friction ($\bar{\mu}^{-1}$), the relevant velocity being that of the equilibrium positions of the beads. The same element $\bar{\Omega}$ appears in the variance. Now, coming to $W$ and $W'$ we have:

$$\langle W \rangle = W_{\text{rev}} + \bar{\Omega}_{AA} v^2 t \tag{65}$$

$$\sigma_W^2 = \frac{2}{\beta} \bar{\Omega}_{AA} v^2 t \tag{66}$$

$$\langle W' \rangle = W_{\text{rev}} - \bar{\Omega}_{AB} v^2 t \tag{67}$$

$$\sigma_{W'}^2 = \frac{2}{\beta} \bar{\Omega}_{BB} v^2 t. \tag{68}$$

Based on these expressions we introduce the asymmetry factor:

$$\text{AF} = \frac{\sigma_W^2 - \sigma_{W'}^2}{\langle W \rangle - \langle W' \rangle}. \tag{69}$$

This quantity relates the distribution of $W$ to $W'$ based on equilibrium information. The AF does not depend either on the pulling speed, which is evident from Eqs. (65),(66),(67),(68) or on the hydrodynamic parameters. An explicit computation gives:

$$\text{AF} = \frac{1}{\beta} \frac{4 k_m (k_A - k_B)}{k_A (k_B + 2 k_m)}. \tag{70}$$

### S6.3 THE AF CAN BE MEASURED FROM EQUILIBRIUM FORCE TRACES

The AF can be directly measured from the equilibrium force traces. The equilibrium distribution for bead positions follows the Boltzmann distribution with respect to the potential in Eq.(39), i.e.

$$P(\mathbf{y}) = Z^{-1} \exp\left(-\frac{\beta}{2} (\mathbf{y} - \mathbf{y}^0) \cdot \bar{\mathsf{k}} (\mathbf{y} - \mathbf{y}^0)\right), \tag{71}$$

According to such distribution variance of $\mathbf{y}$ is:

$$\text{var}(\mathbf{y}) = \beta^{-1} \bar{\mathsf{k}}^{-1}. \tag{72}$$

Moreover, since forces and bead positions are linearly related, $\mathbf{f} = \bar{\mathsf{k}}_D (\mathbf{y} - \mathbf{X}^T)$ we have:

$$\beta \text{var}(\mathbf{f}) = \bar{\mathsf{k}}_D \bar{\mathsf{k}}^{-1} \bar{\mathsf{k}}_D = \frac{1}{k_A k_B + k_m k_A + k_m k_B} \begin{pmatrix} k_A^2 (k_B + k_m) & k_A k_B k_m \\ k_A k_B k_m & k_B^2 (k_A + k_m) \end{pmatrix}, \tag{73}$$

where $\beta = (K_B T)^{-1}$. Using this formula we get:

$$k_A = \beta \text{var}(\mathbf{f})_{AA} + \beta \text{var}(\mathbf{f})_{AB} \tag{74}$$

$$k_B = \beta \text{var}(\mathbf{f})_{BB} + \beta \text{var}(\mathbf{f})_{AB} \tag{75}$$

$$k_m = k_A k_B \frac{\text{var}(\mathbf{f})_{AB}}{\beta \det(\text{var}(\mathbf{f}))}. \tag{76}$$

These formulas assume that the dynamics of the dumbbell is strictly one-dimensional and takes place in the optical plane, i.e. that plane perpendicular to the optical axis. If this is not the case this simplified treatment is not valid anymore and out-of-plane fluctuations must be taken into account. A discussion of these effects can be found in [1]. Once $k_A, k_B$ and $k_m$ are known the AF can be easily computed by Eq. (70). The present method for measuring rigidities from equilibrium force traces was applied and discussed in detail, in a totally different context, in Reference [2]. We used it in this study to extract the values of $k_A, k_B, k_m$ for the asymmetric setup.

## S6.4 DERIVATION OF THE EXPRESSION FOR THE AF

The AF, as given in Eq. (70), can be obtained once the elements of $\bar{\Omega}$ are known. We start from the definition of $\bar{\Omega}$:

$$\bar{\Omega} = \bar{\mathsf{p}}^T \bar{\mu}^{-1} \bar{\mathsf{p}}, \tag{77}$$

where

$$\bar{\mathsf{p}} = \bar{\mathsf{k}}^{-1} \bar{\mathsf{k}}_D = \frac{1}{\mathcal{E}} \left( \begin{array}{cc} k_A(k_B + k_m) & k_B k_m \\ k_A k_m & k_B(k_A + k_m) \end{array} \right), \tag{78}$$

with $\mathcal{E} = k_A k_B + k_A k_m + k_B k_m$. Here it is useful to switch to a representation in which $\bar{\mu}^{-1}$ is diagonal. This can be acheived with a $\pi/4$ rotation:

$$\bar{\mathsf{R}} = \frac{1}{\sqrt{2}} \left( \begin{array}{cc} 1 & -1 \\ 1 & 1 \end{array} \right). \tag{79}$$

Using $\bar{\mathsf{R}}$ we can write:

$$\bar{\Omega} = \bar{\mathsf{p}}^T \bar{\mathsf{R}} \bar{\mathsf{R}}^T \bar{\mu}^{-1} \bar{\mathsf{R}} \bar{\mathsf{R}}^T \bar{\mathsf{p}}, \tag{80}$$

where, as in Section S1, $\bar{\mathsf{R}}^T \bar{\mu}^{-1} \bar{\mathsf{R}}$ is diagonal. The nontrivial contribution which must still be computed is:

$$\bar{\mathsf{p}}' = \bar{\mathsf{R}}^T \bar{\mathsf{p}} = \frac{1}{\sqrt{2}\mathcal{E}} \left( \begin{array}{cc} k_A(k_B + 2k_m) & k_B(k_A + 2k_m) \\ -k_A k_m & k_B k_A \end{array} \right) = \frac{1}{\sqrt{2}\mathcal{E}} \left( \begin{array}{cc} A & B \\ -C & C \end{array} \right). \tag{81}$$

We can thus conclude that:

$$\bar{\Omega} = \frac{1}{2\mathcal{E}^2} \left( \begin{array}{cc} A & -C \\ B & C \end{array} \right) \left( \begin{array}{cc} \gamma_+ A & \gamma_+ B \\ -\gamma_- C & \gamma_- C \end{array} \right) = \frac{1}{2\mathcal{E}^2} \left( \begin{array}{cc} \gamma_+ A^2 + \gamma_- C^2 & \gamma_+ AB - \gamma_- C^2 \\ \gamma_+ AB - \gamma_- C^2 & \gamma_+ B^2 + \gamma_- C^2 \end{array} \right). \tag{82}$$

By definition we have:

$$AF = \frac{2}{\beta} \frac{\bar{\Omega}_{AA} - \bar{\Omega}_{BB}}{\bar{\Omega}_{AA} + \bar{\Omega}_{AB}} = \frac{2}{\beta} \frac{A - B}{A} = \frac{1}{\beta} \frac{4k_m(k_A - k_B)}{k_A(k_B + 2k_m)}. \tag{83}$$

## S6.5 DERIVATION OF THE EQUATIONS

In order to derive the equations of motion (53)-(57) we must apply the Markov Generator to the proper quantities, recalling that

$$\frac{d}{dt} \langle f(\mathbf{y}(t), \mathbf{W}(t)) \rangle_y = \langle L f(\mathbf{y}(t), \mathbf{W}(t)) \rangle_y. \tag{84}$$

We start by deriving Eq.(53). In this case we have that only the first term in the generator Eq.(50) contributes as the second and third terms involve higher derivatives or derivatives with respect to W. We conclude that:

$$\frac{d}{dt} \langle \mathbf{y}(t) \rangle_y = \langle L \mathbf{y}(t) \rangle_y = -\bar{\mu} \bar{\mathsf{k}} (\langle \mathbf{y}(t) \rangle_y - \mathbf{y}^0(t)). \tag{85}$$

In the same way we can derive the equation for $\bar{\sigma}_y^2$, it is useful to recall that

$$\langle (\mathbf{y} - \langle \mathbf{y} \rangle_y) \otimes (\mathbf{y} - \langle \mathbf{y} \rangle_y) \rangle_y = \langle \mathbf{y} \otimes \mathbf{y} \rangle_y - \langle \mathbf{y} \rangle_y \otimes \langle \mathbf{y} \rangle_y, \tag{86}$$

so that

$$\frac{d}{dt} \bar{\sigma}_y^2 = \langle L (\mathbf{y} \otimes \mathbf{y}) \rangle_y - \frac{d}{dt} \langle \mathbf{y} \rangle_y \otimes \langle \mathbf{y} \rangle_y \tag{87}$$

we will work out the two terms of the right hand side of (87) separately, starting from the first:

$$\langle L (\mathbf{y} \otimes \mathbf{y}) \rangle_y = \langle \left( -\bar{\mu} \bar{\mathsf{k}} (\mathbf{y} - \mathbf{y}^0) \cdot \nabla_y + \beta^{-1} \bar{\mu} \cdot \nabla_y \nabla_y \right) \mathbf{y} \otimes \mathbf{y} \rangle_y. \tag{88}$$

The third term in the generator was neglected because it involves a derivative with respect to $\mathbf{W}$. Acting with the derivatives on $\mathbf{y} \otimes \mathbf{y} \rangle_y$ gives:

$$\begin{aligned} \langle L (\mathbf{y} \otimes \mathbf{y}) \rangle_y &= -\langle \bar{\mu} \bar{\mathsf{k}} \left( (\mathbf{y} - \mathbf{y}^0) \otimes \mathbf{y} \right) - (\mathbf{y} \otimes (\mathbf{y} - \mathbf{y}_0)) (\bar{\mu} \bar{\mathsf{k}})^T \rangle_y + 2\beta^{-1} \bar{\mu} \\ &= -2 \mathrm{Sym} \left( \langle \bar{\mu} \bar{\mathsf{k}} \left( (\mathbf{y} - \mathbf{y}^0) \otimes \mathbf{y} \right) \rangle_y \right) + 2\beta^{-1} \bar{\mu}. \end{aligned} \tag{89}$$

The second term in (87) is just a time derivative, which can be worked out thanks to (85):

$$\frac{d}{dt}\langle \mathbf{y}\rangle_y \otimes \langle \mathbf{y}\rangle_y = -\bar{\mu}\bar{\mathsf{k}}\left((\langle \mathbf{y}\rangle_y - \mathbf{y}^0) \otimes \langle \mathbf{y}\rangle_y\right) - \left(\langle \mathbf{y}\rangle_y \otimes (\langle \mathbf{y}\rangle_y - \mathbf{y}^0)\right)(\bar{\mu}\bar{\mathsf{k}})^T =$$
$$= -2\mathrm{Sym}\left(\bar{\mu}\bar{\mathsf{k}}\left((\langle \mathbf{y}\rangle_y - \mathbf{y}^0) \otimes \mathbf{y}\right)\right). \tag{90}$$

An explicit result fot the time derivative of $\bar{\sigma}_y^2$ is obtained by taking the difference of (89) and (90)

$$\frac{d}{dt}\bar{\sigma}_y^2 = -2\mathrm{Sym}\left(\langle \bar{\mu}\bar{\mathsf{k}}\left((\mathbf{y} - \mathbf{y}^0) \otimes \mathbf{y}\right)\rangle_y - \bar{\mu}\bar{\mathsf{k}}\left((\langle \mathbf{y}\rangle_y - \mathbf{y}^0) \otimes \langle \mathbf{y}\rangle_y\right)\right) + 2\beta^{-1}\bar{\mu} =$$
$$= -2\mathrm{Sym}\left(\langle \bar{\mu}\bar{\mathsf{k}}\left(\mathbf{y} \otimes \mathbf{y}\right)\rangle_y - \bar{\mu}\bar{\mathsf{k}}\left(\langle \mathbf{y}\rangle_y \otimes \langle \mathbf{y}\rangle_y\right)\right) + 2\beta^{-1}\bar{\mu} =$$
$$= -2\mathrm{Sym}\left(\bar{\mu}\bar{\mathsf{k}}\bar{\sigma}_y^2\right) + 2\beta^{-1}\bar{\mu}. \tag{91}$$

In the second step above we used the fact that

$$\langle\left((\mathbf{y} - \mathbf{y}^0) \otimes \mathbf{y}\right)\rangle_y = \langle(\mathbf{y} \otimes \mathbf{y})\rangle_y - \left(\mathbf{y}^0 \otimes \langle \mathbf{y}\rangle_y\right)$$

while

$$\left((\langle \mathbf{y}\rangle_y - \mathbf{y}^0) \otimes \langle \mathbf{y}\rangle_y\right) = \left(\langle \mathbf{y}\rangle_y \otimes \langle \mathbf{y}\rangle_y\right) - \left(\mathbf{y}^0 \otimes \langle \mathbf{y}\rangle_y\right).$$

Following the order of equations (53)-(57) we should now compute the average of $\mathbf{W}$. This is again a simple matter, as the computation proceeds almost exactly as in the case of $\langle \mathbf{y}\rangle_y$, with the only difference that now only the last term in Eq.(50) for the generator matters. The result is:

$$\frac{d}{dt}\langle \mathbf{W}\rangle_y = v\bar{\mathsf{k}}_D(\langle \mathbf{y}\rangle_y - \mathbf{X}^T(t)). \tag{92}$$

The case of $\bar{\mathsf{C}}$, the cross-correlation of $\mathbf{y}$ and $\mathbf{W}$ involves some additional computations. The function $\bar{\mathsf{C}}$ is linear in both $\mathbf{W}$ and $\mathbf{y}$ so that only the first and third terms in the generator contribute. The time derivative of $\bar{\mathsf{C}} = (\mathbf{W} - \langle \mathbf{W}\rangle_y) \otimes (\mathbf{y} - \langle \mathbf{y}\rangle_y)$ again yields two pieces, as in the case of (87), and we shall proceed in a similar way:

$$\frac{d}{dt}\bar{\mathsf{C}} = \langle L(\mathbf{W} \otimes \mathbf{y})\rangle_y - \frac{d}{dt}\langle W\rangle_y \otimes \langle \mathbf{y}\rangle_y. \tag{93}$$

Again we will compute the two parts separately:

$$\langle L(\mathbf{W} \otimes \mathbf{y})\rangle_y = \langle -\left(\mathbf{W} \otimes (\bar{\mu}\bar{\mathsf{k}}(\mathbf{y} - \mathbf{y}^0))\right) + \left((v\bar{\mathsf{k}}_D(\mathbf{y} - \mathbf{X}^T)) \otimes \mathbf{y}\right)\rangle_y =$$
$$= \langle -\bar{\mu}\bar{\mathsf{k}}\left(\mathbf{W} \otimes (\mathbf{y} - \mathbf{y}^0)\right) + v\left((\mathbf{y} - \mathbf{X}^T) \otimes \mathbf{y}\right)\bar{\mathsf{k}}_D\rangle_y, \tag{94}$$

and

$$\frac{d}{dt}\langle \mathbf{W}\rangle_y \otimes \langle \mathbf{y}\rangle_y = -\bar{\mu}\bar{\mathsf{k}}\left(\langle \mathbf{W}\rangle_y \otimes (\langle \mathbf{y}\rangle_y - \mathbf{y}^0)\right) + v\left((\langle \mathbf{y}\rangle_y - \mathbf{y}^0) \otimes \langle \mathbf{y}\rangle_y\right)\bar{\mathsf{k}}_D. \tag{95}$$

Taking the difference of these two terms and using

$$\langle\left(\mathbf{W} \otimes (\mathbf{y} - \mathbf{y}^0)\right)\rangle_y = \langle(\mathbf{W} \otimes \mathbf{y})\rangle_y - \left(\langle \mathbf{W}\rangle_y \otimes \mathbf{y}^0\right)$$

and

$$\left(\langle \mathbf{W}\rangle_y \otimes (\langle \mathbf{y}\rangle_y - \mathbf{y}^0)\right) = \left(\langle \mathbf{W}\rangle_y \otimes \langle \mathbf{y}\rangle_y\right) - \left(\langle \mathbf{W}\rangle_y \otimes \mathbf{y}^0\right)$$

it is possible to conclude that:

$$\frac{d}{dt}\bar{\mathsf{C}} = -\bar{\mu}\bar{\mathsf{k}}\langle(\mathbf{W} - \langle \mathbf{W}\rangle_y) \otimes (\mathbf{y} - \langle \mathbf{y}\rangle_y)\rangle_y + \langle(\mathbf{y} - \langle \mathbf{y}\rangle_y) \otimes (\mathbf{y} - \langle \mathbf{y}\rangle_y)\rangle_y\bar{\mathsf{k}}_D$$
$$= -\bar{\mu}\bar{\mathsf{k}}\bar{\mathsf{C}} + v\bar{\sigma}_y^2\bar{\mathsf{k}}_D. \tag{96}$$

The only equation left to derive is that for $\bar{\sigma}_W^2 = \langle(\mathbf{W} - \langle \mathbf{W}\rangle_y) \otimes (\mathbf{W} - \langle \mathbf{W}\rangle_y)\rangle_y$. This computation will only involve the last term in the generator:

$$\frac{d}{dt}\langle(\mathbf{W} - \langle \mathbf{W}\rangle_y) \otimes (\mathbf{W} - \langle \mathbf{W}\rangle_y)\rangle_y = \langle L(\mathbf{W} \otimes \mathbf{W})\rangle_y - \frac{d}{dt}\left(\langle \mathbf{W}\rangle_y \otimes \langle \mathbf{W}\rangle_y\right). \tag{97}$$

As usual we will evaluate the two terms separately and put them together as a second step:

$$\langle L\left(\mathbf{W}\otimes\mathbf{W}\right)\rangle_y = v\langle\bar{\mathsf{k}}_D\left(\mathbf{W}\otimes(\mathbf{y}-\mathbf{X}^T)\right) + \left((\mathbf{y}-\mathbf{X}^T)\otimes\mathbf{W}\right)\bar{\mathsf{k}}_D\rangle_y,\tag{98}$$

$$\frac{d}{dt}\left(\langle\mathbf{W}\rangle_y\otimes\langle\mathbf{W}\rangle_y\right) = v\bar{\mathsf{k}}_D\left(\mathbf{W}\otimes(\langle\mathbf{y}\rangle_y-\mathbf{X}^T)\right) + \left((\langle\mathbf{y}\rangle_y-\mathbf{X}^T)\otimes\mathbf{W}\right)\bar{\mathsf{k}}_D.\tag{99}$$

The last equation of motion is thus:

$$\frac{d}{dt}\bar{\sigma}_W^2 = +v\bar{\mathsf{k}}_D\langle(\mathbf{y}-\langle\mathbf{y}\rangle_y)\otimes(\mathbf{W}-\langle\mathbf{W}\rangle_y)\rangle_y + v\langle(\mathbf{W}-\langle\mathbf{W}\rangle_y)\otimes(\mathbf{y}-\langle\mathbf{y}\rangle_y)\rangle_y\bar{\mathsf{k}}_D =$$
$$= +2\mathrm{Sym}\left(v\bar{\mathsf{k}}_D\bar{\mathsf{C}}\right).\tag{100}$$

## S6.6 SOLUTION OF THE EQUATIONS

In the previous section we introduced an average operation $\langle\cdot\rangle_y$ without specifying with respect to which measure the average is taken. In this section we will suppose that the initial probability distribution for the process is $p(\mathbf{y},0) = \delta(\mathbf{y}-y)$, a deterministic initial condition. The Work is by definition equal to 0 at $t=0$, again a deterministic initial condion, and the same is true for the different variances and covariances beacuse of the initial conditions. The initial conditions for the equations of motion will then be:

$$\begin{cases}\langle\mathbf{y}(0)\rangle_y = y\\ \langle\mathbf{W}(0)\rangle_y = \bar{\sigma}_{\mathbf{y}}^2(0) = \bar{\mathsf{C}}(0) = \bar{\sigma}_{\mathbf{W}}^2(0) = 0.\end{cases}\tag{101}$$

All the equations we have to solve are linear equations which can either be solved by variation of constants or by direct integration. The first equation, that for $\langle\mathbf{y}\rangle_y$ is a case for variation of constants:

$$\begin{cases}\langle\mathbf{y}(0)\rangle_y = y\\ \frac{d}{dt}\langle\mathbf{y}\rangle_y = -\bar{\mu}\bar{\mathsf{k}}(\langle\mathbf{y}\rangle_y-\mathbf{y}^0(t)).\end{cases}\tag{102}$$

The solution is

$$\langle\mathbf{y}(t)\rangle_y = e^{-\bar{\mu}\bar{\mathsf{k}}t}\left(y+\int_0^t\bar{\mu}\bar{\mathsf{k}}e^{\bar{\mu}\bar{\mathsf{k}}s}\mathbf{y}^0(s)ds\right),\tag{103}$$

as it can be directly checked. We recall the definition of $\mathbf{y}^0(t)$

$$\mathbf{y}^0(t) = \left(\lambda+vt-\frac{k_Ak_m}{\det(\bar{\mathsf{k}})}(\lambda_0+vt),\frac{k_Bk_m}{\det(\bar{\mathsf{k}})}(\lambda_0+vt)\right).\tag{104}$$

In the following it will be useful to write $\mathbf{y}^0(t) = \psi+\xi t$ where

$$\psi = \left(\lambda_0-\frac{k_Ak_m}{\det(\bar{\mathsf{k}})}\lambda_0,\frac{k_Bk_m}{\det(\bar{\mathsf{k}})}\lambda_0\right)\tag{105}$$

$$\xi = \left(v-\frac{k_Ak_m}{\det(\bar{\mathsf{k}})}v,\frac{k_Bk_m}{\det(\bar{\mathsf{k}})}v\right).\tag{106}$$

In terms of these quantities (103) reads:

$$\langle\mathbf{y}(t)\rangle_y = e^{-\bar{\mu}\bar{\mathsf{k}}t}\left(y+\int_0^t\bar{\mu}\bar{\mathsf{k}}e^{\bar{\mu}\bar{\mathsf{k}}s}(\psi+\xi t)ds\right).\tag{107}$$

Integrating by parts we get:

$$
\begin{aligned}
\langle \mathbf{y}(t) \rangle_y &= e^{-\bar{\mu}\bar{\mathsf{k}}t} \left( y + \int_0^t \bar{\mu}\bar{\mathsf{k}} e^{\bar{\mu}\bar{\mathsf{k}}s}(\psi + \xi t) ds \right) = \\
&= e^{-\bar{\mu}\bar{\mathsf{k}}t} \left( y + e^{\bar{\mu}\bar{\mathsf{k}}s}(\psi + \xi s)\big|_0^t - \int_0^t e^{\bar{\mu}\bar{\mathsf{k}}s}\xi ds \right) = \\
&= e^{-\bar{\mu}\bar{\mathsf{k}}t} \left( y + e^{\bar{\mu}\bar{\mathsf{k}}t}(\psi + \xi s) - \psi - (\bar{\mu}\bar{\mathsf{k}})^{-1} e^{\bar{\mu}\bar{\mathsf{k}}s}\xi\big|_0^t \right) = \\
&= e^{-\bar{\mu}\bar{\mathsf{k}}t} \left( y + e^{\bar{\mu}\bar{\mathsf{k}}t}(\psi + \xi t) - (\bar{\mu}\bar{\mathsf{k}})^{-1} \left( e^{\bar{\mu}\bar{\mathsf{k}}t} - 1 \right) \xi \right) = \\
&= \left( e^{-\bar{\mu}\bar{\mathsf{k}}t}(y - \psi) + (\psi + \xi t) - (\bar{\mu}\bar{\mathsf{k}})^{-1} \left( 1 - e^{-\bar{\mu}\bar{\mathsf{k}}t} \right) \xi \right) = \\
&= e^{-\bar{\mu}\bar{\mathsf{k}}t}(y - \psi) + \mathbf{y}^0(t) - (\bar{\mu}\bar{\mathsf{k}})^{-1} \left( 1 - e^{-\bar{\mu}\bar{\mathsf{k}}t} \right) \xi
\end{aligned}
\tag{108}
$$

The average $\langle \mathbf{W} \rangle_y$ is given by:

$$
\begin{aligned}
\langle \mathbf{W} \rangle_y &= \bar{\mathsf{k}}_D \int_0^t \langle \mathbf{y}(s) \rangle_y - \mathbf{X}^T(s) ds = \\
&= \bar{\mathsf{k}}_D \int_0^t e^{-\bar{\mu}\bar{\mathsf{k}}s}(y - \psi) + \mathbf{y}^0(s) - (\bar{\mu}\bar{\mathsf{k}})^{-1} \left( 1 - e^{-\bar{\mu}\bar{\mathsf{k}}s} \right) \xi - \mathbf{X}^T(s) ds = \\
&= \bar{\mathsf{k}}_D \left( \int_0^t \left( \mathbf{y}^0(s) - \mathbf{X}^T(s) ds \right) - (\bar{\mu}\bar{\mathsf{k}})^{-1} e^{-\bar{\mu}\bar{\mathsf{k}}s}(y - \psi)\big|_0^t \right. \\
&\quad \left. - (\bar{\mu}\bar{\mathsf{k}})^{-1} \left( s + (\bar{\mu}\bar{\mathsf{k}})^{-1} e^{-\bar{\mu}\bar{\mathsf{k}}s} \right)\big|_0^t \xi \right) = \\
&= \bar{\mathsf{k}}_D \left( \int_0^t \left( \mathbf{y}^0(s) - \mathbf{X}^T(s) ds \right) - (\bar{\mu}\bar{\mathsf{k}})^{-1} \left( e^{-\bar{\mu}\bar{\mathsf{k}}s} - 1 \right)(y - \psi) \right. \\
&\quad \left. + (\bar{\mu}\bar{\mathsf{k}})^{-1} \left( t + (\bar{\mu}\bar{\mathsf{k}})^{-1} \left( e^{-\bar{\mu}\bar{\mathsf{k}}t} - 1 \right) \right) \xi \right)
\end{aligned}
\tag{109}
$$

The equation for the variance $\bar{\sigma}_y^2$ requires some more attention, we first prove a useful identity: let $A$ and $B$ be symmetric matrices. The product $AB$ is not in general symmetric

$$
(AB)^T = BA.
$$

As a consequence the exponential of $AB$, $e^{AB}$ is not symmetric. We want now to prove that:

$$
\left( A^{-1} e^{AB} \right) = \left( A^{-1} e^{AB} \right)^T = e^{BA} A^{-1}.
\tag{110}
$$

Using the definition of the matrix exponential we have:

$$
\begin{aligned}
A^{-1} e^{AB} &= A^{-1} \sum_{k=0}^{\infty} \frac{(AB)^k}{k!} = A^{-1} + B + BAB + \cdots = (1 + BA + BABA + \dots) A^{-1} \\
&= \sum_{i=0}^{\infty} \frac{(BA)^i}{i!} A^{-1} = e^{BA} A^{-1}.
\end{aligned}
\tag{111}
$$

Similarly one can prove:

$$
\left( A e^{AB} \right) = \left( A e^{AB} \right)^T = e^{BA} A.
\tag{112}
$$

These results will prove useful in soving the equation for $\bar{\sigma}_y^2$. We will first solve:

$$
\begin{cases}
\bar{\sigma}_{\mathbf{y}}^2(0) = 0 \\
\frac{d}{dt} \bar{\sigma}_y^2 = -2\bar{\mu}\bar{\mathsf{k}}\bar{\sigma}_y^2 + 2\frac{\bar{\mu}}{\bar{\beta}},
\end{cases}
\tag{113}
$$

and then prove that the solution of (113) also solves (54). Equation (113) can again be solved by variation of the constants:

$$\bar{\sigma}_{\mathbf{y}}^2(t) = e^{-2\bar{\mu}\bar{k}t} \int_0^t e^{2\bar{\mu}\bar{k}s} 2\frac{\bar{\mu}}{\bar{\beta}} ds = e^{-2\bar{\mu}\bar{k}t} \left( e^{2\bar{\mu}\bar{k}s}(\bar{\mu}\bar{k})^{-1}\frac{\bar{\mu}}{\bar{\beta}} \right)\Big|_0^t =$$
$$= \left( 1 - e^{-2\bar{\mu}\bar{k}t} \right) \frac{\bar{k}^{-1}}{\bar{\beta}}. \tag{114}$$

The original equation for $\bar{\sigma}_y^2$ contains $\mathrm{Sym}\left(\bar{\mu}\bar{k}\bar{\sigma}_y^2\right)$. Using (110),(112) we can prove that:

$$\bar{\mu}\bar{k}\bar{\sigma}_{\mathbf{y}}^2(t) = \bar{\mu}\bar{k} \left( 1 - e^{-2\bar{\mu}\bar{k}t} \right) \frac{\bar{k}^{-1}}{\bar{\beta}} = \frac{\bar{k}^{-1}}{\bar{\beta}} \left( 1 - e^{-2\bar{\mu}\bar{k}t} \right)^T k\mu = \left( \bar{\mu}\bar{k}\bar{\sigma}_{\mathbf{y}}^2(t) \right)^T. \tag{115}$$

In other words

$$\bar{\mu}\bar{k}\bar{\sigma}_{\mathbf{y}}^2(t) = \mathrm{Sym}\left(\bar{\mu}\bar{k}\bar{\sigma}_{\mathbf{y}}^2(t)\right),$$

so that if $\bar{\sigma}_{\mathbf{y}}^2(t)$ solves (113) it also solves (54). The equation for $\bar{\mathsf{C}}$, is again solved by variation of constants:

$$\begin{cases} \bar{\mathsf{C}}(0) = 0 \\ \frac{d}{dt}\bar{\mathsf{C}} = -\bar{\mu}\bar{k}\bar{\mathsf{C}} + v\bar{\sigma}_y^2\bar{k}_D. \end{cases} \tag{116}$$

The solution is:

$$\bar{\mathsf{C}}(t) = v e^{-\bar{\mu}\bar{k}t} \int_0^t e^{\bar{\mu}\bar{k}s}\bar{\sigma}_{\mathbf{y}}^2(s)\bar{k}_D ds =$$
$$= v e^{-\bar{\mu}\bar{k}t} \int_0^t e^{\bar{\mu}\bar{k}s} \left( 1 - e^{-2\bar{\mu}\bar{k}s} \right) \frac{\bar{k}^{-1}\bar{k}_D}{\beta} =$$
$$= v e^{-\bar{\mu}\bar{k}t} \int_0^t \left( e^{\bar{\mu}\bar{k}s} - e^{-\bar{\mu}\bar{k}s} \right) \frac{\bar{k}^{-1}\bar{k}_D}{\beta} =$$
$$= v e^{-\bar{\mu}\bar{k}t}(\bar{\mu}\bar{k})^{-1} \left( e^{\bar{\mu}\bar{k}s} + e^{-\bar{\mu}\bar{k}s} \right)\Big|_0^t \frac{\bar{k}^{-1}\bar{k}_D}{\beta} = \tag{117}$$
$$= v e^{-\bar{\mu}\bar{k}t}(\bar{\mu}\bar{k})^{-1} \left( e^{\bar{\mu}\bar{k}t} + e^{-\bar{\mu}\bar{k}t} - 2 \right) \frac{\bar{k}^{-1}\bar{k}_D}{\beta} =$$
$$= v(\bar{\mu}\bar{k})^{-1} \left( 1 + e^{-2\bar{\mu}\bar{k}t} - 2e^{-\bar{\mu}\bar{k}t} \right) \frac{\bar{k}^{-1}\bar{k}_D}{\beta} =$$
$$= v(\bar{\mu}\bar{k})^{-1} \left( 1 - e^{-\bar{\mu}\bar{k}t} \right)^2 \frac{\bar{k}^{-1}\bar{k}_D}{\beta}.$$

As in the case of $\langle W \rangle_y$ the equation for $\bar{\sigma}_W^2$ is solved by direct integration:

$$\bar{\sigma}_W^2 = \int_0^t 2v^2\bar{k}_D\bar{\mathsf{C}}(s)ds = \int_0^t v\bar{k}_D(\bar{\mu}\bar{k})^{-1} \left( 1 - e^{-\bar{\mu}\bar{k}s} \right)^2 \frac{\bar{k}^{-1}\bar{k}_D}{\beta} =$$
$$= 2v^2 \int_0^t \bar{k}_D(\bar{\mu}\bar{k})^{-1} \left( 1 - 2e^{-\bar{\mu}\bar{k}s} + e^{-2\bar{\mu}\bar{k}s} \right) \frac{\bar{k}^{-1}\bar{k}_D}{\beta} =$$
$$= 2v^2\bar{k}_D(\bar{\mu}\bar{k})^{-1} \left( s + 2(\bar{\mu}\bar{k})^{-1}e^{-\bar{\mu}\bar{k}s} - \frac{(\bar{\mu}\bar{k})^{-1}}{2}e^{-2\bar{\mu}\bar{k}s} \right)\Big|_0^t \frac{\bar{k}^{-1}\bar{k}_D}{\beta} = \tag{118}$$
$$= 2v^2\bar{k}_D(\bar{\mu}\bar{k})^{-1} \left( t + 2(\bar{\mu}\bar{k})^{-1} \left( e^{-\bar{\mu}\bar{k}t} - 1 \right) - \frac{(\bar{\mu}\bar{k})^{-1}}{2} \left( e^{-2\bar{\mu}\bar{k}s} - 1 \right) \right) \frac{\bar{k}^{-1}\bar{k}_D}{\beta} =$$
$$= v^2\bar{k}_D(\bar{\mu}\bar{k})^{-2} \left( 2\bar{\mu}\bar{k}t + 2 \left( e^{-\bar{\mu}\bar{k}t} - 1 \right) - \left( e^{-\bar{\mu}\bar{k}s} - 1 \right)^2 \right) \frac{\bar{k}^{-1}\bar{k}_D}{\beta}$$

Keeping terms of order $\mathcal{O}(t)$ in Eqs. (108) and (118) we get the expressions anticipated in Eqs. (58) and (59)

# S7  EXPERIMENTS AT LOWER PULLING SPEED

In the main text we present experiments performed at high pulling speed, in order to enhance the dissipation associated with the movement of the center-of-mass. This was done to highlight the fact that $W'$ does not fulfill the CFR. Nevertheless such violation is still evident at lower pulling speeds. Just fpr completeness in figure S7 we show the distributions of $W_-, W, W'$ and $W_+$ in a pulling experiment performed on a dsDNA tether (experimental conditions are identical to those reported in the main paper) at a pulling speed of 500 nm/s with $\Delta\lambda = 400$ nm.

## CAPTION FOR FIGURE 1

**Experimental set-up.** Two laser beams, oriented along the $z$ direction, are used to create two optical traps. A dumbbell is formed by two optically trapped beads and a molecular tether. The tether is oriented along the $y$ direction, perpendicular to the optical axis $z$. We choose the center of one trap (trap B) as the origin of our coordinate system. $\lambda$ denotes the trap–to–trap distance while $y_A$ and $y_B$ denote the positions of the centers of the beads with respect to the reference trap B.

## CAPTION FOR FIGURE 2

**Independent coordinates in linear systems** A) Two dimensional histograms of the forces measured in a **symmetric** dual trap setup. The dashed line forms a $\pi/4$ angle with the coordinate axes and corresponds to the definition of $f_+$. B) Covariance $\langle f_+^\phi f_-^\phi \rangle$ as a function of $\phi$ in a symmetric setup. The red line denotes 0 covariance (i.e. linear independence). At the bottom of the graph we report the result of a 1% significance $\chi^2$ test, with red meaning dependent and green meaning independent. C) Two dimensional histograms of the force as measured in an **asymmetric** dual trap setup. The dashed line forms a $\pi/4$ angle with the coordinate axes and corresponds to the definition of $f_+$. In this asymmetric case $f_+, f_-$ do not correspond to the principal axes of the histogram. D) Covariance $\langle f_+^\phi f_-^\phi \rangle$ as a function of $\phi$ in a **asymmetric** setup. The red line denotes 0 covariance (i.e. linear independence). It is still possible to define a coordinate system where the two degrees of freedom are uncoupled, but now the definition is different from that given in the main text. At the bottom of the graph we report the result of a 1% significance $\chi^2$ test, with red meaning dependent and green meaning independent. Tests were performed on 3 seconds data traces with 1kHz acquisition rate.

## CAPTION FOR FIGURE 3

**Independent coordinates in non-linear systems** Left panels: force fluctuations of a two-state DNA hairpin. Force increases from top to bottom and the hairpin is 1) completely folded, 2) preferentially folded, 3) preferentially unfolded, 4) completely unfolded. Right panels: Covariance $\langle f_+^\phi f_-^\phi \rangle$ as a function of $\phi$, The red line denotes 0 covariance (i.e. linear independence). At the bottom of the graphs we report

the result of a 1% significance $\chi^2$ test, with red meaning dependent and green meaning independent. Data show that the center of mass ($x_+$) and differential coordinate ($x_-$) are independent. Tests were performed on 3 seconds data traces with 1kHz acquisition rate.

### CAPTION FOR FIGURE 4

**Direct measurements $\gamma$ and $\Gamma$.** Left panel: hydrodynamic friction ($\gamma$, solid symbols) and interaction coefficient ($\Gamma$, open symbols) measured from the decay rate of thermal fluctuations. Each point is the average upon measurements over 5 different molecules. Solid lines are the theoretical predictions Eqs. (31),(32). The horizontal dashed line marks the exact theoretical value at contact ($\rho = 2$). Right panel: Ratio between the hydrodynamic coefficients ($\Gamma/\gamma$) as a function of $(\rho - 2)^{-1}$. This quantity does not depend directly on $r_b$. Open symbols represent measurements obtained with untethered beads at different separations, solid symbols show measurements obtained with tethered beads. The continuous line gives the theoretical prediction according to the first two term in the expansion as in Eqs. (31),(32).

### CAPTION FOR FIGURE 5

**Convolution and reconstruction of $P(W)$ from the CFR for the asymmetric case.** Each row shows the result of a convolution by fixing the value of $\Sigma$ and changing $\Delta$. The top, middle and bottom rows correspond to $\Sigma^2 = 0, \Sigma^2 = 2.5, \Sigma^2 = 7.5$. In each panel we show $P_{\Delta,\Sigma}(W)$ (blue solid points), together with $P_{\Delta,\Sigma}(-W)\exp(W)$ (blue open points) and the experimentally measured $P(W)$ (green points). Three different convolutions are found to fulfill the fluctuation symmetry $P_{\Delta,\Sigma}(W) = P_{\Delta,\Sigma}(-W)\exp(W)$, (panels in the rightmost column), but only one of them (bottom right panel) is compatible with the AF and matches the true $P(W)$.

### CAPTION FOR FIGURE 6

**Asymmetry factor and inference in asymmetric setups** Left panel: The asymmetry factor as a function of $k_A/k_B$ and $k_m/k_B$. Different curves correspond to different values of $k_m/k_B$ (0.3,0.5,1,2,3 from bottom to top). The dashed curve corresponds to the limit $k_m = \infty$. In the symmetric case ($k_A = k_B = 1$) the different curves coincide ($\Sigma^2 = 0$). The red point denotes the asymmetric conditions in which experiments were performed. Right panel: Inference in the asymmetric case. Different pairs ($\Delta, \Sigma^2$) yield probability distributions satisfying the CFR (blue points and blue line). The asymmetry factor (red line) selects a narrow range of possible values, which is compatible with the experimentally measured values $\Delta = 8.2\ K_B\mathrm{T}$ and $\Sigma^2 = 7.2\ (K_B\mathrm{T})^2$.

### CAPTION FOR FIGURE 7

**Work measurements at lower pulling speed.** The statistics of $W, W', W_+$ and $W_-$ are shown. The pulling speed $v = 500$ nm/s is less then half of the lowest pulling speed presented in the main text, but the effect on the validity of the CFR for the different work quantities ($W, W', W^+, W^-$) is still visible. 500 nm/s lies in the typical range of pulling speed used in single molecule pulling experiments.

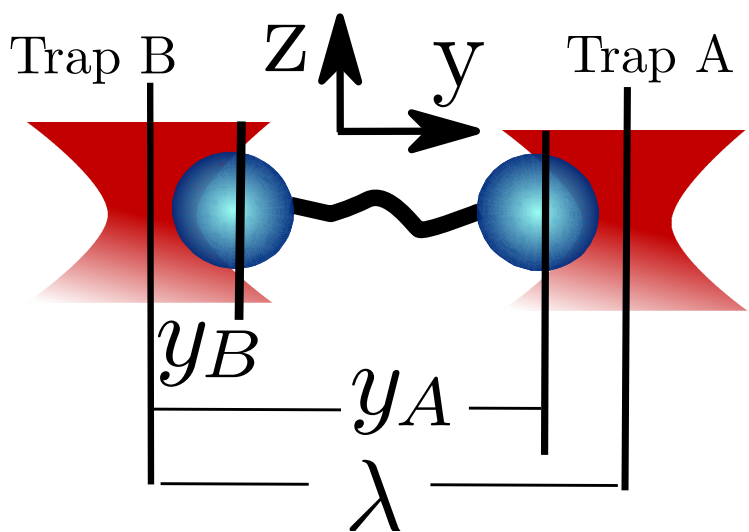

Figure S 1

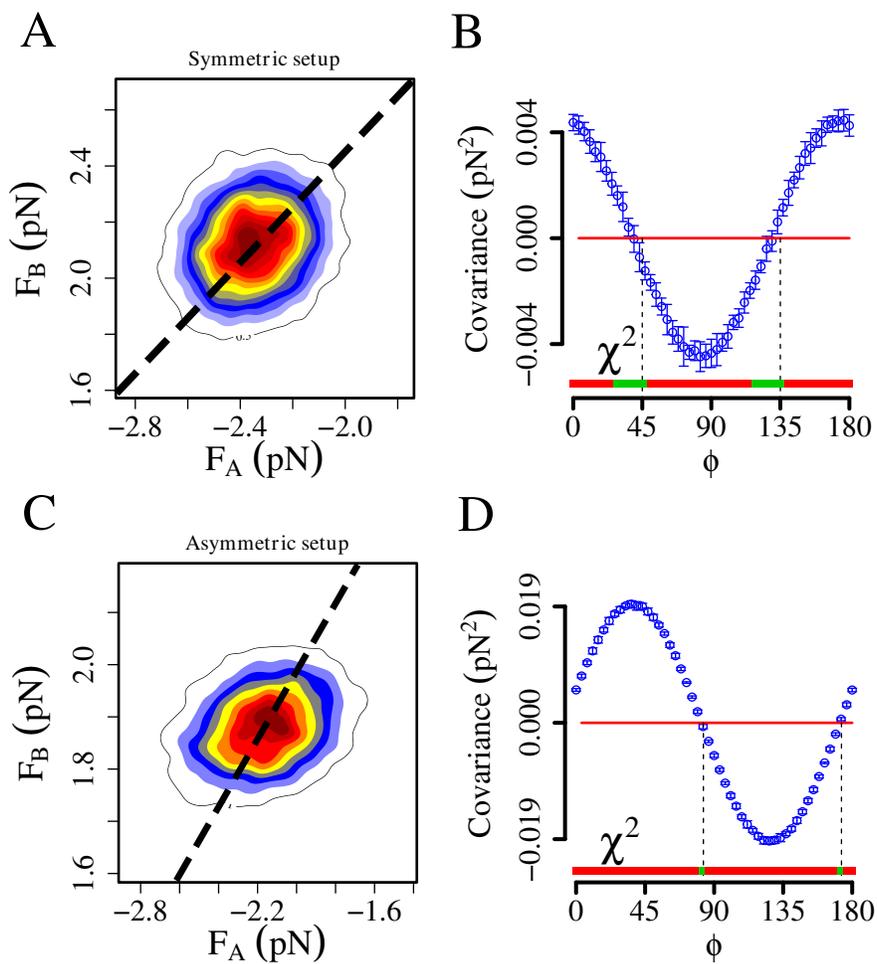

Figure S 2

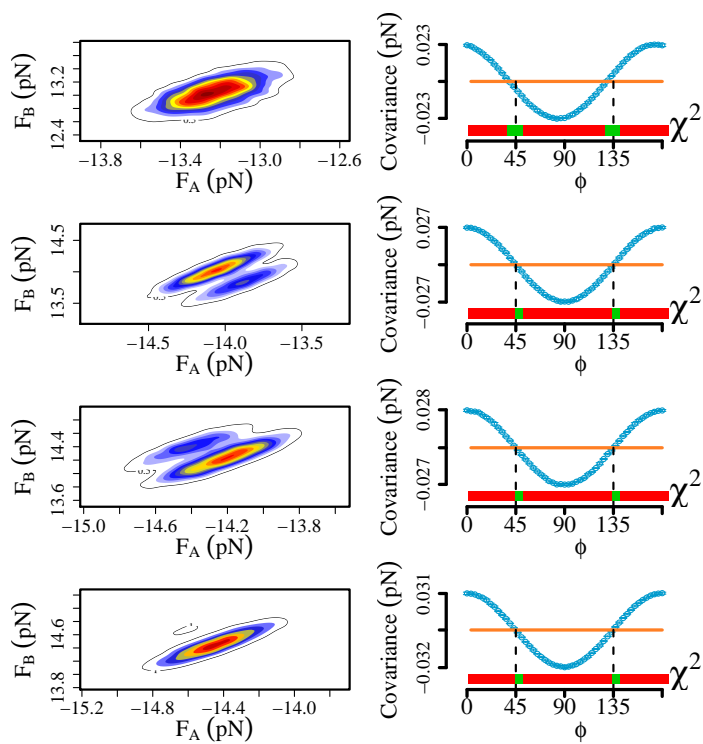

Figure S 3

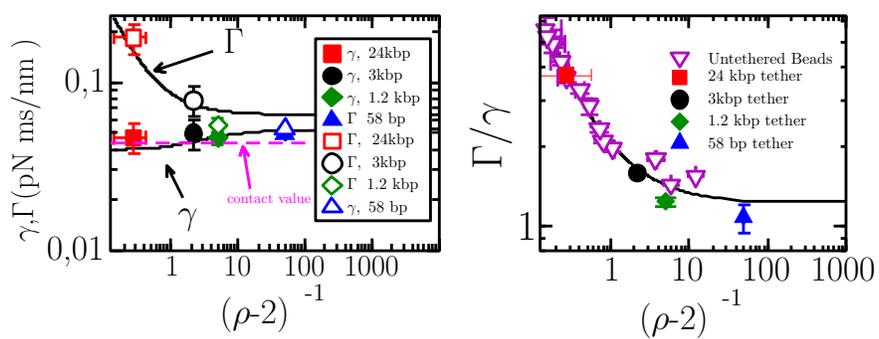

Figure S 4

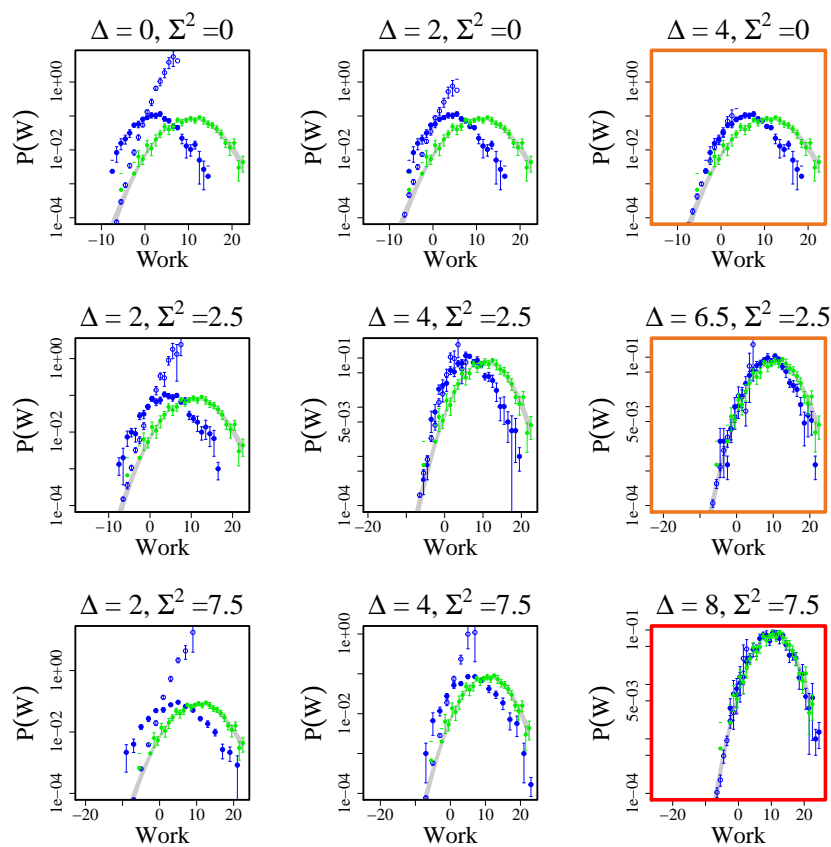

Figure S 5

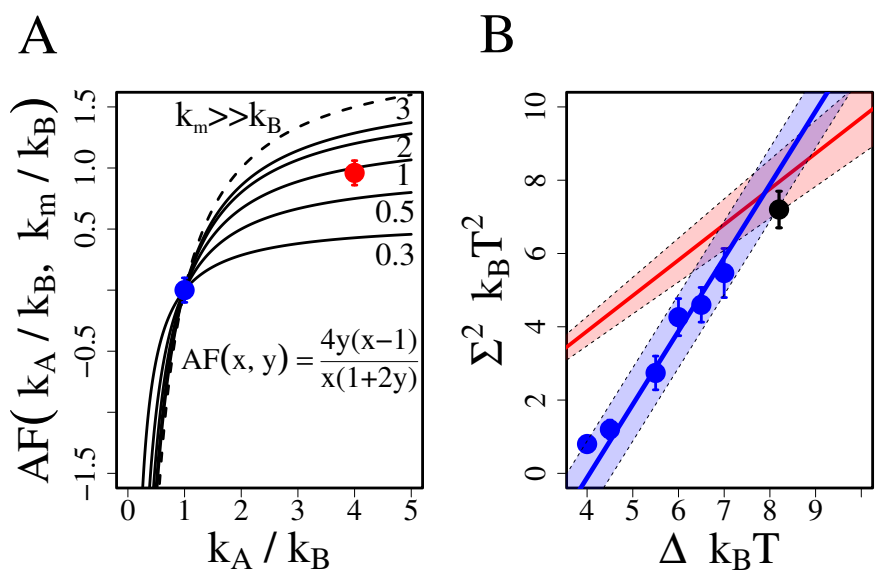

Figure S 6

v=500 nm/s

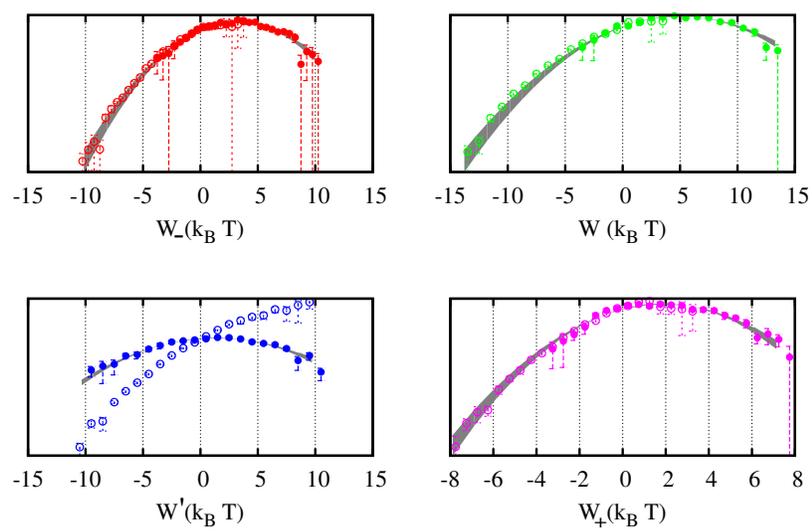

Figure S 7

TABLE S1

Table S 1: Molecule 1

| Pulling speed | cov($W$,$W'$) | $\chi^2$-test[1] | cov($W^+$,$W^-$) | $\chi^2$-test[1] |
|---|---|---|---|---|
| | | Forward protocol | | |
| 7.2 $\mu$m/s | 0.41 | 1 | 0.08 | 0 |
| 4.3 $\mu$m/s | 0.35 | 1 | 0.09 | 0 |
| 1.35 $\mu$m/s | 0.38 | 1 | 0.05 | 0 |
| | | Reverse protocol | | |
| 7.2 $\mu$m/s | 0.17 | 0 | 0.018 | 0 |
| 4.3 $\mu$m/s | -0.05 | 0 | -0.05 | 0 |
| 1.35 $\mu$m/s | 0.36 | 1 | 0.20 | 0 |
| | | Cyclic protocol | | |
| 7.2 $\mu$m/s | 0.12 | 1 | 0.00 | 0 |
| 4.3 $\mu$m/s | 0.12 | 1 | 0.01 | 0 |
| 1.35 $\mu$m/s | 0.34 | 1 | 0.09 | 0 |

[1] 1=Dependent 0=Independent

TABLE S2

Table S 2: Molecule 2

| Pulling speed | cov($W$,$W'$) | $\chi^2$-test[1] | cov($W^+$,$W^-$) | $\chi^2$-test[1] |
|---|---|---|---|---|
| | | Forward protocol | | |
| 7.2 $\mu$m/s | 0.469377 | 1 | -0.12 | 1 |
| 4.3 $\mu$m/s | 0.6170601 | 1 | -0.14 | 0 |
| 1.35 $\mu$m/s | 0.5346567 | 1 | 0.09 | 0 |
| | | Reverse protocol | | |
| 7.2 $\mu$m/s | -0.31 | 1 | 0.25 | 0 |
| 4.3 $\mu$m/s | -0.37 | 1 | 0.20 | 0 |
| 1.35 $\mu$m/s | -0.73 | 1 | 0.19 | 0 |
| | | Cyclic protocol | | |
| 7.2 $\mu$m/s | 0.26 | 1 | -0.19 | 1 |
| 4.3 $\mu$m/s | 0.34 | 1 | -0.18 | 0 |
| 1.35 $\mu$m/s | 0.42 | 1 | -0.10 | 0 |

[1] 1=Dependent 0=Independent



\end{document}